%% file: 0-main.tex
\documentclass[sigconf]{aamas} 

\usepackage{balance} 
\usepackage{multirow}
\usepackage{multicol}

\usepackage{diagbox}
\usepackage{graphicx}
\usepackage{subcaption}

\usepackage{amsfonts,amssymb, amsmath} 
\usepackage{bm} 

\usepackage{xcolor} 
\newcommand{\jqruan}[1]{\textcolor{magenta}{jqruan:[#1]}}

\newcommand{\yali}[1]{\textcolor{red}{[yali:#1]}}

\usepackage{changes}
\usepackage{enumitem}
\usepackage{float}

\newcommand{\namegg}{graph generator}
\newcommand{\namefp}{graph-based coordinated policy}
\newcommand{\namefcp}{Graph-based Coordinated Policy}

\newtheorem{proposition}{Proposition}

\newtheorem{definition}{Definition}
\usepackage[ruled]{algorithm2e} 

\setcopyright{ifaamas}
\acmConference[AAMAS '22]{Proc.\@ of the 21st International Conference
on Autonomous Agents and Multiagent Systems (AAMAS 2022)}{May 9--13, 2022}
{Online}{P.~Faliszewski, V.~Mascardi, C.~Pelachaud,
M.E.~Taylor (eds.)}
\copyrightyear{2022}
\acmYear{2022}
\acmDOI{}
\acmPrice{}
\acmISBN{}

\acmSubmissionID{337}

\title{GCS: Graph-Based Coordination Strategy for Multi-Agent Reinforcement Learning}

\author{Jingqing Ruan$^{12}$, 
Yali Du$^{\ast 3}$, 
Xuantang Xiong$^{1}$, 
Dengpeng Xing$^{1}$, 
Xiyun Li$^{1}$, 
Linghui Meng$^{1}$, 
Haifeng Zhang$^{1}$, 
Jun Wang$^{4}$,
Bo Xu$^{\ast 1}$}
\thanks{$\ast$ Corresponding author}
\affiliation{
  \institution{
  $^1$Institute of Automation, Chinese Academy of Sciences, Beijing, China\\
  $^2$School of Future Technology, University of Chinese Academy of Sciences, Beijing, China\\
  $^3$King's College London,  \ \  $^4$University College London
  }
  \city{}
  \country{}
}
\email{{ruanjingqing2019,xiongxuantang2021,lixiyun2020,dengpeng.xing,menglinghui2019,haifeng.zhang,xubo}@ia.ac.cn}
\email{yali.du@kcl.ac.uk,  jun.wang@cs.ucl.ac.uk}

\begin{abstract}

Many real-world scenarios involve a team of agents that have to coordinate their policies to achieve a shared goal.
Previous studies mainly focus on decentralized control to maximize a common reward and barely consider the coordination among control policies, which is critical in dynamic and complicated environments.
In this work, we propose factorizing the joint team policy into a~\namegg~and~\namefp~to enable coordinated behaviours among agents.
The~\namegg~adopts an encoder-decoder framework that outputs directed acyclic graphs (DAGs) to  capture the underlying dynamic decision structure.
We also apply the DAGness-constrained and DAG depth-constrained optimization in the graph generator to balance efficiency and performance.
The~\namefp~exploits the generated decision structure.
The graph generator and coordinated policy are trained simultaneously to maximize the discounted return. 
Empirical evaluations on Collaborative Gaussian Squeeze, Cooperative Navigation, and Google Research Football demonstrate the superiority of the proposed method.
The code is available at \url{https://github.com/Amanda-1997/GCS_aamas337}.

\end{abstract}


\newcommand{\BibTeX}{\rm B\kern-.05em{\sc i\kern-.025em b}\kern-.08em\TeX}

\begin{document}


\pagestyle{fancy}
\fancyhead{}


\maketitle

\input{1-introduction}
\input{2-related}
\input{2-problem-setup}

\input{3-method}

\input{4-experiment}
\input{4-exp-performance}
\input{4-exp-hierachy}

\input{4-exp-robust}

\input{5-conclusion}
\input{6-ack}




\newpage
%

\bibliographystyle{unsrt}
\bibliography{sample}

\appendix
\include{appendix}

\end{document}

%% file: 1-introduction.tex
\section{Introduction}

Multi-agent reinforcement learning (MARL) has shown exceptional results in many real-life applications, such as multiplayer games~\cite{vinyals2017starcraft,lowe2017maddpg}, traffic control~\cite{kuyer2008traffic}, and social dilemmas~\cite{leibo2017multi_social}. 
A suitable control policy is extremely important in multi-agent systems (MASs). 
One choice is to treat the MAS as a single agent and adopt a centralized control policy~\cite{han2019grid,jiang2018atoc}; however, this approach is constrained by  poor scalability for high-dimensional state and action spaces.
On the contrary, decentralized control ~\cite{sunehag2018vdn,rashid2018qmix,du2019liir,lowe2017maddpg,iqbal2019maac} allows agents to make decisions independently, but struggles to enable coordinated behaviors on complex tasks.
Taking traffic flow as an example, when multiple vehicles are trying to cross an intersection without traffic lights, most likely, the traffic will become congested if all vehicles take actions simultaneously without a rational sequence.
This problem may be solved, however, if those vehicles move in an orderly way based on some coordination structure. 
This example shows that it is imperative to improve on a fully decentralized decision-making process, and a solution to alleviate the above issue is to develop a coordinated control policy to obtain cooperative behaviors.

Several approaches have been reported to address the problem of action coordination.
BiAC~\cite{zhang2020bilevel} mainly focuses on  coordination of the asynchronous decisions of two agents.
The multi-agent rollout algorithm~\cite{bertsekas2019multiagent_rollout} provides a theoretical view of executing a local rollout with some coordinating information, but is limited to an agent-by-agent decision dependency structure.
Although these works investigate the action execution order of two-agent and multi-agent systems, they are still insufficient to characterize the complicated and dynamic underlying decision dependency structure of a general multi-agent system. 
Moreover, we assert that representing the underlying decision dependency structure and using this to control the action execution is essential to improving coordination.


In this work, we propose a
graph-based coordination strategy (GCS)
that learns coordinated behaviors through factorizing the joint team policy into a~\namegg~and a~\namefp. 
The former aims to learn an action coordination graph (ACG) that properly represents the decision dependency.
The latter further coordinates the dependent behaviors among agents exploiting the underlying decision dependency.
We train the~\namegg~and the~\namefp~simultaneously to maximize the discounted return.
For the ACG we employ directed acyclic graphs (DAGs),
whose nodes represent agents and whose directed edges denote action dependencies of the associated agents. 
Moreover, we propose using the DAGness-constrained and DAG depth-constrained optimization in the~\namegg~to balance efficiency and performance.

\vspace{4pt}
The contributions of this paper can be summarized as follows:
\begin{itemize}[itemsep= 4 pt,topsep = 6 pt]
    \item As far as we know, we are the first to introduce  directed acyclic graphs to action coordination, dynamically representing the underlying decision dependencies of MAS.
    \item We propose a DAGness-constrained and DAG depth-constrained optimization in the~\namegg, achieving a trade-off between decision-making efficiency and performance.
    \item Empirical evaluations on several challenging MARL benchmarks (Collaborative Gaussian Squeeze, Cooperative Navigation, and Google Football) show that our method can achieve superior performance and obtain meaningful results consistent with intuitive expectations.
\end{itemize}

%% file: 2-related.tex
\section{Related Work}



Deep reinforcement learning has been successfully applied to addressing complex decision problems ~\citep{silver2017alphagozero,schrittwieser2020muzero,zhang2021population,wang2021ordering,meng2021offline}. 
Due to the widespread existence of multi-agent tasks, MARL has attracted increasing attention, and learning appropriate control policies is important to obtain the maximum cumulative discounted return.
Based on the structures of their execution schemes, 
we classify the existing approaches into three categories.

First, a fully independent execution scheme allows agents to determine actions according to their individual policies without any interaction.
One line of research, such as IQL~\cite{tan1993iql}, VDN~\citep{sunehag2018vdn}, QMIX~\citep{rashid2018qmix}, and QTRAN~\citep{son2019qtran}, focuses on value-based methods, which assign each agent an independent policy for execution.
Another line of research, including MAAC~\citep{iqbal2019maac}, COMA~\citep{foerster2018coma}, and LIIR~\citep{du2019liir}, extends the actor-critic algorithm~\cite{degris2012actorcritic} to the multi-agent case, where each actor represents an individual policy for an agent. 

Second, the communication-based independent execution scheme is widely used, which allows the use of extensive information in its individual decision making~\citep{busoniu2008comprehensive_survey}. In this scheme, agents learn how to transmit informative messages and to process the messages during training. Then agents exchange the messages to determine their actions individually during independent execution. 
Representative methods~\cite{foerster2016RIALDIAL,sukhbaatar2016CommNet,zhang2013coordinating,peng2017bicnet,jiang2018atoc,du2021flowcomm} autonomously learn communication protocols that are required in generating informative messages: these determine whom to communicate with and what messages to transmit for assisted decision making. 

Third, the coordinated execution scheme, where agents develop their policies conditioned on other agents' actions and make decisions in a coordinated manner, is important in MAS.
There are some methods that implicitly model the coordinated behaviours from the perspective of a coordination graph. DCG~\cite{bohmer2020dcg} uses pairwise graphs to propagate beliefs for joint decisions, while DICG~\cite{li2021dicg} focuses on generating a coordination graph with soft edge weights for message passing. DGN~\cite{jiang2018dgn} uses a graph attention network as an embedding extractor to assist in the decision making.
Furthermore, some methods have been proposed to explicitly model the coordinated behaviours in order, such as Bi-AC~\citep{zhang2020bilevel} and multi-agent rollout~\citep{bertsekas2019multiagent_rollout}, which  propose  utilizing two-agent and agent-by-agent dependency structures, respectively, to help agents make decisions in order and promote action coordination. 
Similar but essentially distinct, we introduce a mechanism to learn the underlying DAG structure that represents the decision dependency among agents.

Moreover, the generation of the DAGs is an essential part of our work. Recently, some continuous optimization approaches~\cite{zheng2018dagsnotears,yu2019daggnn,yu2020dagsnocurl,lachapelle2020gradientdag} have been proposed to recover the DAGs through structure learning. The method~\cite{zhu2019dagRL}, closely related to our work, uses reinforcement learning as its search strategy to maximize a predefined score function. 
Borrowing from an idea of~\cite{zhu2019dagRL}, which obtains the DAG using reinforcement learning, we construct a graph generator module to generate the DAG structure as an action coordination graph and regard the extrinsic reward as the incentives to jointly train with MARL tasks.

%% file: 2-problem-setup.tex
\section{Problem Setup}

\paragraph{\textbf{MMDPs} }
Cooperative multi-agent problems can be modeled as multiagent Markov decision processes (MMDPs)~\cite{boutilier1996planning}, which can be expressed as a tuple $<\{\mathcal{I}\}, \mathcal{S},\{\mathcal{U}^i\}_{i=1}^N, \mathbb{P},r, \gamma>$.
$i \in \mathcal{I}$ is the $i^{th}$ player, 
$\mathcal{S}$ is the global state space, 
and $\mathcal{U}^i$ denotes the action space for the $i^{th}$ player.
We label $\bm{u}:=(u^1,...,u^N)$ the joint actions for all players. 
Intuitively, the agent $i$ will select an individual action $u^i$ to perform and execute it.
$\mathbb{P}$ is the transition dynamics, and $\mathbb{P}(\cdot |s,\bm{u})$ gives the distribution of the next state at state $s$ taking action $\bm{u}$. 
All agents share the same reward function $r(s,\bm{u}):s \times \bm{u} \to R$. 
$\gamma \in (0,1)$
denotes a discount factor, and $\tau  = ({s_0},{\bm{u}_0},{s_1},...)$ 
denotes the trajectory induced by the policy $\bm{\pi}=\{\pi^i\}_{i=1}^N$. All the agents coordinate
together to maximize the cumulative discounted return ${\mathbb{E}_{\bm{\tau}  \sim \pi }}\left[ {\sum\nolimits_{t = 0}^\infty  {{\gamma ^t}r({{s}_t},{\bm{u}_t})} } \right]$.

\paragraph{\textbf{ Factored MMDPs} }
We formalize our problem based on MMDPs as factored MMDPs.
Different from MMDPs, where all actions are taken simultaneously and do not depend on each other, we endow the hierarchy order to the joint action based on the learned DAG structure $\mathcal{G}$, called an action coordination graph (ACG).
The adjacency matrix $A$ representing the ACG denotes the decision dependency from the~\namegg~$\rho$.
With $A$, we can define $ \bm{\pi} = \prod\limits_{i = 1}^n {\pi ^{{i}}}({u^i}|{o^i},{u^{pa(i)}})$ as the~\namefp~for the $i^{th}$ player, where $o^i$ is the observation of the i-th player, $pa(i)$ are the parents of agent $i$, and ${u^{pa(i)}}$ are the actions taken by the parents, whose order is generated from $A$. 
Note that a fully decentralized policy is a special case of our~\namefp~where none of the agents have  parents. 

\begin{figure}[t!]
    \centering
      \includegraphics[width=0.9\linewidth]{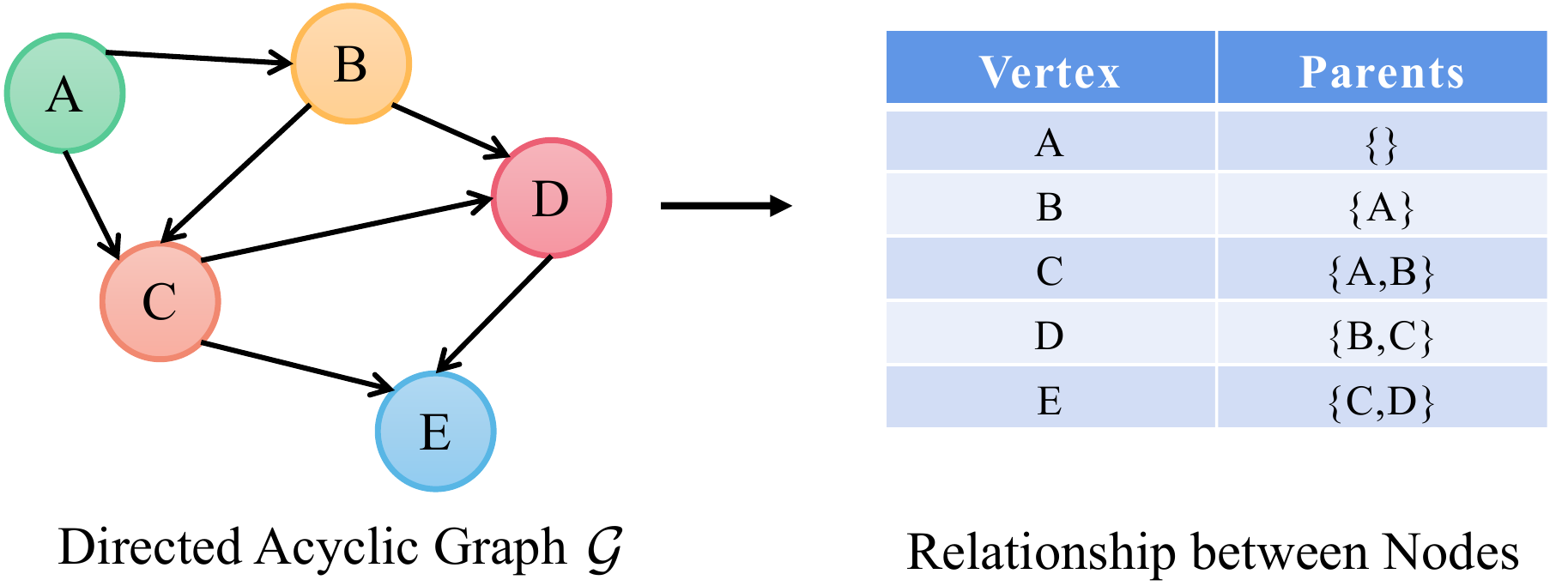}
\caption{The left panel is a schematic diagram of a directed acyclic graph $\mathcal{G}$ with vertices $|\mathcal{V}|=5$ and edges $|\mathcal{E}|=7$. The right panel shows the parent--child relationships of $\mathcal{G}$.}
	  \label{fig:demo-topo}
\end{figure}

Figure \ref{fig:demo-topo} gives an example of the DAG and the relationships between nodes.
The nodes in $\mathcal{G}$ correspond to the agents in the MAS, and the parent-child relationships represent the hierarchical decision dependencies among agents.
For example, the case that the node $D$ in the graph has two parents $\{B,C\}$ illustrates that the action taken by the agent $D$ is constrained by $B$ and $C$.

Now, the graph-based coordination strategy is factorized as: 
\begin{equation}
\label{eq:factor}
   \pi (\bm{u}, A|s)  = \rho ( A|s)\prod\limits_{i = 1}^n {{\pi ^i}({u^i}|{o^i},{u^{pa(i) \sim A}})},
\end{equation}
where $\rho$ is the~ \namegg, and $\pi^i$ is~\namefp. 



%% file: 3-method.tex
\section{Methodology}





The overall goal is to maximize the cumulative return, denoted as:
\begin{equation}
\label{eq:J}
\eta  = {\mathbb{E}_{A\sim \rho ,\bm{u}\sim \pi ( \cdot |s,A)}}\left[ {\sum\limits_{k = 0}^\infty  {{\gamma ^k}r({s_{t + k}},{\bm{u}_{t + k}})} } \right].
\end{equation}
Now we further elaborate on the derivation of the~\namefp~${\pi _{{\theta ^i}}}({u^i}|{o^i},{u^{pa(i)}})$ and the~\namegg~ $\rho_\varphi(A|s)$, respectively. 


\begin{figure*}[ht!]
\begin{center}
\includegraphics[width=0.9\linewidth]{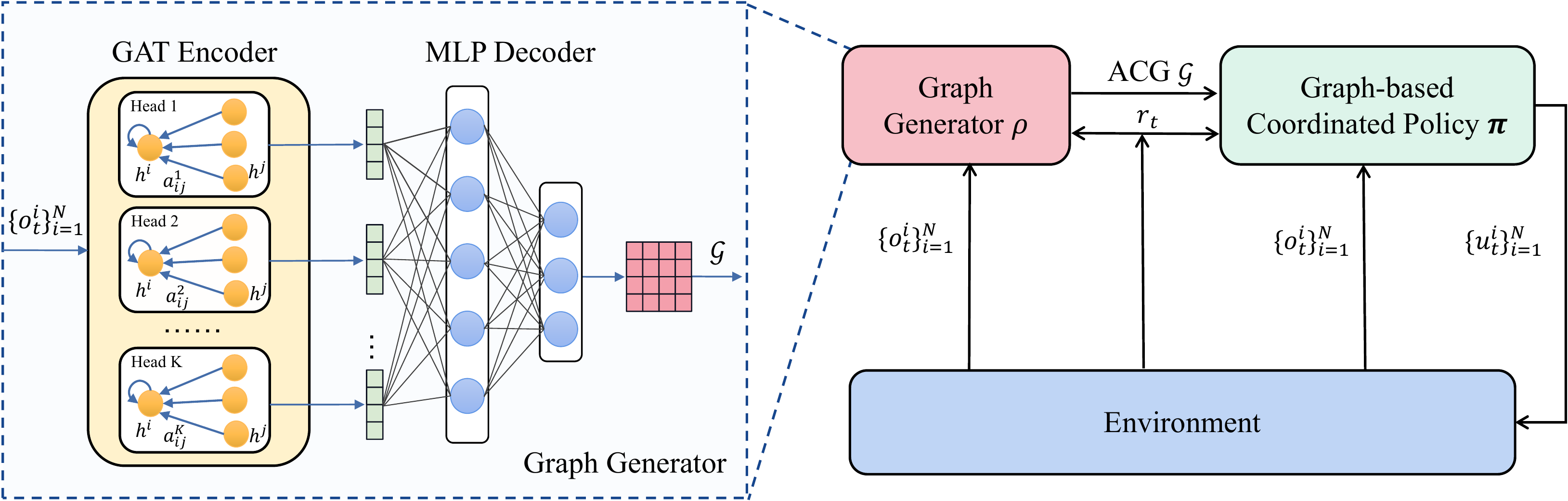}
\end{center}
\caption{The proposed framework for the interaction process. The left is the graph generator including an encoder-decoder neural network model which is used to generate the DAG from the observed data. }
\label{fig:overview}
\end{figure*}

\subsection{\namefcp}
Given a known~\namegg~$\rho$ (elaborated in Section~\ref{sec:acg}), we have $A \sim \rho$ to represent the underlying decision dependency. Based on it, we can denote the decision policy as $\bm{\pi} = \prod\limits_{i = 1}^N {{\pi ^i}({u^i}|{o^i},{u^{pa(i)\sim A}})}$, called~\namefp.
We explore~\namefp~ $\bm{\pi}^*$ that obtains the final joint action $\{{u}^i\}_{i=1}^N$ as follows.

With Equations~(\ref{eq:factor}) and (\ref{eq:J}), we can write the expected return for agent $i$ as:
\begin{equation}
    \begin{array}{l}
\eta^i = {\mathbb{E}_{s\sim {p^{\bm{\pi}} },(\bm{u},A)\sim  \pi }}\left[ { \pi (\bm{u},A|s)Q_{\bm{\pi}} ^i(s,\bm{u})} \right]\\
 = {\mathbb{E}_{s\sim {p^{\bm{\pi}} },{u^i}\sim {\pi ^i},A\sim \rho }}\left[ {\rho (A|s)\prod\limits_{i = 1}^N {{\pi ^i}({u^i}|{o^i},{u^{pa(i)\sim A}})} Q_{\bm{\pi}} ^i(s,\bm{u})} \right].
\end{array}
\end{equation}

The~\namefp~$\bm{\pi}=\{\pi_1, \pi_2,...,\pi_N\}$ for $N$ agents can be parameterized by $\bm{\theta}=\{\theta_1, \theta_2,...,\theta_N\}$.
Correspondingly, the gradient of the expected return for agent $i$ is expressed as:
\begin{equation}
\label{eq:policy}
    \begin{array}{l}
{\nabla _{{\theta ^i}}}\eta({\theta ^i}) = {\nabla _{{\theta ^i}}}{\eta ^i} \vspace{2pt} \\
 = {\mathbb{E}_{s\sim {p^{\bm{\pi}} },u^i \sim \pi^i }}\left[ { {\nabla _{{\theta ^i}}}\log{\pi ^i}({u^i}|{o^i},{u^{pa(i)}}) \sum\nolimits_A {\rho (A|s)}  Q_{\bm{\pi}} ^i(s,\bm{u})} \right].
\end{array}
\end{equation}

By applying the mini-batch technique to 
the off-policy training, the gradient can be approximately estimated as:
\begin{equation}
    {\nabla _{{\theta ^i}}}\eta({\theta ^i}) = {\mathbb{E}_{(s,\bm{o}, A, {\bm{u}})\sim \mathcal{D}}}\left[ { {\nabla _{{\theta ^i}}}\log{\pi ^i}({u^i}|{o^i},{u^{pa(i)}})Q_{\bm{\pi}} ^i(s,{\bm{u}})} \right],
\end{equation}
where $\mathcal{D}$ is the experience replay buffer, recording experiences of all agents.
Moreover, the centralized action-value function $Q_\pi ^i$ can be updated as:
\begin{equation}
\label{eq:critic}
{\mathcal{L}}(\phi ) = {E_{(s,\bm{o}, A, {\bm{u}})\sim \mathcal{D}}}\left[ {{{\left( {Q_{\bm{\pi}} ^i(s,\bm{u}) - {y^i}} \right)}^2}} \right]
\end{equation}
where ${y^i} = r + \gamma {\max _{\bm{u}'}}Q_{{{\bm{\pi}} ^ - }}^i\left( {s',\bm{u}'} \right)$ is the learning target 
and $Q_{{{\bm{\pi}} ^ - }}^i$ is the target network parameterized by $\theta^{i-}$.

During the training process, the~\namefp~ $\bm{\pi}=\{\pi_1, \pi_2,...,\pi_N\}$ and the ~\namegg~$\rho$ are updated iteratively. We will describe how to find the~\namegg~$\rho(\cdot)$ under a given policy $\bm{\pi}$.
\subsection{Graph Generator}
\label{sec:acg}
The~\namegg~aims to generate the DAG $\mathcal{G}$ to define the decision dependency among agents. 
We will introduce it in detail from three aspects: (a) DAGness constraint, (b) DAG Depth constraint, and (c) optimization objective.

\paragraph{\textbf{DAGness constraint}}
The acyclicity constraint is an important issue in our problem setting. In this work, we also use the penalty terms like~\cite{zheng2018dagsnotears,lachapelle2020gradientdag,zhu2019dagRL} to ensure acyclicity. The result in~\cite{zheng2018dagsnotears} shows that the directed graph $\mathcal{G}$ with binary adjacency matrix $A$ is acyclic if and only if:
\begin{equation}
\label{eq:dag_constr}
    g(A):=\operatorname{trace}(e^{A \circ A})-d=0,
\end{equation}
where $e^{A \circ A}$ is the matrix exponential, $A \circ A$ guarantees the non-negativity, and $d$ is the number of nodes in the DAGs.
The ‘$\operatorname{trace}$’ of a matrix is defined as the sum of the diagonal elements~\cite{zheng2018dagsnotears}.
The constraint function $g$ should satisfy that: (a) its derivatives are computable, and (b) $g$ can be the measurement of DAGs. 

\paragraph{\textbf{DAG Depth constraint}}

Moreover, taking the trade-off between efficiency and performance into account, we claim that the maximum depth of graph structure should be adjustable over  different tasks. 
Therefore, we propose an alternative constraint to control the hierarchy of the generated DAGs as follows.

\begin{definition}
A square matrix $A$ is a Nilpotent Matrix~\cite{Algebra75}, if
\begin{equation*}
    {A^k} = O \ and \ {A^{k - 1}} \ne O, \ \exists k \in {\mathop{\rm \mathbb{Z^+}}} ,
\end{equation*}
where $O$ is the zero matrix and $A$ is called the Nilpotent of index $k$.
\end{definition}
\begin{proposition}
\label{pro:hierarchy}
Let $A$ be an adjacency matrix for a directed acyclic graph, then the maximal length between any two nodes $i$ and $j$ is $k$ if $A$ is Nilpotent of index $k$.
\end{proposition}

We provide a detailed proof of  
proposition~\ref{pro:hierarchy} in Appendix~\ref{proof:hierarchy}. 
Here, we define ${c}(A^k): = sum({A^k}) = \sum\nolimits_i {\sum\nolimits_j {A_{ij}^k = 0} }$, which is equivalent to $A^k = O$. 
We remark that as long as constraint ${A^k} = O$ holds, it can be guaranteed that the maximal length between any two nodes $i$ and $j$ of the DAG does not exceed $k$.





\paragraph{\textbf{Optimization objective}.} 
Based on the foregoing, we can optimize $\rho$ parameterized by $\varphi$ with the maximal length $k$ by:
\begin{equation}
\label{eq:cons_1}
    \begin{array}{l}
\max \ \eta({\varphi}) = {\mathbb{E} }\left[ {\sum\limits_{k = 0}^\infty  {{\gamma ^k}r^{\rho}({s_{t + k}},{\bm{u}_{t + k}})} } \right] \vspace{2pt} \\
s.t. \ g(W) = 0,{c}(W^k) = 0
\end{array},
\end{equation}
where $W = \rho_{\varphi}(\cdot)$ denotes the weight matrix generated from the~\namegg~$\rho$.
Then the weight matrix is modeled as a Bernoulli distribution, from which the binary adjacency matrix $A$ is sampled. Here, we use the constraints of the weight matrix
$g(W)$ and $c(W^k)$ 
to approximate those of the adjacency matrix $g(A)$ and $c(A^k)$ due to the consistency of representing the graph structure. With this approximation, we restate $\eta^i$ as: 
\begin{equation}
  {\eta ^i} = \mathbb{E}\left[ {\rho \left( {W|s} \right)\prod\limits_{i = 1}^N {{\pi ^i}\left( {{u^i}|{o^i},{u^{pa(i)}}} \right)Q_{\bm{\pi}} ^i(s,\bm{u})} } \right].  
\end{equation}


Fixing~\namefp~$\bm{\pi}$, we approximate the~\namegg~$\rho$ as follows.
We augment the original problem shown in Equation~(\ref{eq:cons_1}) with a quadratic penalty using the augmented Lagrangian technique~\cite{bertsekas1997nonlinear}:
\begin{equation}
\label{eq:aug}
  \begin{array}{l}
\mathop {\min }  \limits_\varphi  \quad - \eta (\varphi ) + \frac{\xi }{2}\left[ {||g(W)||^2 + ||c( {{W^k}} )||^2} \right]\\
s.t. \quad  g(W) = 0,c( {{W^k}} ) = 0
\end{array},
\end{equation}
with the penalty $\xi>0$.

Next, we convert the Equation~(\ref{eq:aug}) to an  unconstrained Lagrangian function:
\begin{equation}
\label{eq:aug_lag}
    \begin{array}{l}
L(\varphi ,{\lambda }_1,{\lambda}_2) =  - \eta ({\varphi }) + \frac{\xi }{2}\left[ {||g({W })||^2 + ||c\left( {{{{W}}^k}} \right)||^2} \right]\\
 \qquad \qquad \qquad  + {\lambda }_1g({W }) + {\lambda }_2c\left( {{{{W}}^k}} \right)\\
\end{array}.
\end{equation}

\begin{proposition}
The gradient for Equation~(\ref{eq:aug_lag}) to optimize the coordination graph generation policy can be derived as follows:
%
\label{pro:grad_rho}
\begin{equation*}
\begin{array}{l}
\begin{array}{*{20}{l}}
{{\nabla _\varphi }L({\varphi}, {\lambda}_1, {\lambda}_2 ) = {E_{s \sim p,W \sim \rho_{\varphi} ( \cdot |s)}}\left[ {{\nabla _\varphi }\log \rho_{\varphi} (W|s)\sum\limits_{{u^i}} {{{ \pi }^i}({u^i}|{o^i}, } } \right.} \vspace{2px} \\
{\qquad \qquad {u^{pa(i)}}) \cdot
{Q{{^i}}}({o^i},{u^i})
- {{\lambda} _1}{{({e^{W^\circ W}})}^T} \cdot 2W - }
\end{array} \vspace{2px} \\
\qquad \qquad \left. {{{\lambda} _2}{{\sum\nolimits_{i,j} {\left[ {k{W^k}{W^{ - 1}}} \right]} }_{ij}}} \right],
\end{array}
\end{equation*}

\end{proposition}

We provide a detailed proof of proposition~\ref{pro:grad_rho} in Appendix~\ref{proof:grad_rho}.

In proposition~\ref{pro:grad_rho}, 
we remark that after considering the influence of various decision dependencies on the reinforcement learning tasks, we can obtain the underlying graph structure that makes the best response to MARL tasks.


\subsection{Implementation Details}
As shown in~Figure~\ref{fig:overview}, the proposed framework include the~\namefp~and the~\namegg, which will be elaborated below.
\paragraph{\textbf{\namefcp}}
The~\namefp~can be obtained from the standard multi-agent actor-critic framework.
As for the policy $\pi^i$ of the actor, we use the RNN network with the stochastic policy gradient to model the action distributions. The critic used to criticize the  joint actions made by the actors is a three-layer feed-forward neural network activated by the ReLU units, denoted as $f_{critic}(o_t,u_t)=ReLU(\textit{MLP}\langle o_t,u_t \rangle)$. 




\paragraph{\textbf{Graph {Generator}}}
As shown in Figure~\ref{fig:overview}, the graph generator $\rho_{\varphi}(\bm{o})$ adopt an encoder-decoder module used to find the ACG. 
The GAT-based encoder can model the interplay of agents and extract the further latent representations.
The MLP-based decoder is used to recover the pairwise relationship between agents to generate an ACG.
The graph generator takes the local observations of the agents as input and outputs the ACG $\mathcal{G}$ to obtain the decision dependency for the decision-making process of the~\namefp, elaborated as follows.

The graph generator contains two sub-modules. Firstly, we use the graph attention network (GAT)~\cite{velivckovic2018gat} as the attention-based encoder to extract the latent information for the graph structure generation. First, the simple feature is extracted by a multi-layer perceptron (MLP) as an initial step:
\begin{equation}
    {\{h_t^i\}}_{i=1}^N = f_{MLP}({\{o_t^i\}}_{i=1}^N, {\{u_{t-1}^i\}}_{i=1}^N).
\end{equation}

Due to the sufficiently expressive power of GAT, we use it to extract 
the further latent information of the simple feature.
We compute the importance coefficients through the attention mechanism:
\begin{equation}
    a_t^{ij} = \frac{{\exp (h_t^i\bm{W}{{(h_t^j\bm{W})}^T}/\sqrt {{d}} ))}}{{\sum\nolimits_{k \in {{\{ I\} }^{ - i}}} {\exp (h_t^i\bm{W}{{(h_t^k\bm{W})}^T}/\sqrt {{d}} )} }},\sum\nolimits_{j \in {{\{ I\} }^{ - i}}} {a_t^{ij}}  = 1,
\end{equation}
where $\bm{W} \in R^{d{'} \times d}$ is a learnable weight matrix, $d{'}$ and $d$ denote the dimensions of the input vector and the latent vector, respectively, and
$k \in {{\{ I\} }^{-i}}$ indexes other agents except the agent $i$.

Then the multi-head attention is used to stabilize the learning process of self-attention, and the final latent feature is as follows:
\begin{equation}
    h_t^{i'} = \sigma \left( {\frac{1}{M}\sum\limits_{m = 1}^M {\sum\limits_{j \in {{\{ I\} }^{ - i}}} {a_{t,m}^{ij}{\bm{W}^m}h_t^j} } } \right).
\end{equation}
where $M$ is the number of attention heads. $a_{t,m}^{ij}$ are importance coefficients computed by the k-th attention mechanism, and $W^m$ is the  corresponding weight matrix.

Another sub-module in the graph generator
is the decoder that generates 
a weight matrix used to sample the graph structure.
Since the GAT-based encoder has already provided sufficiently expressive features among agents, a single-layer decoder is enough to easily construct the pairwise relationship between the encoder outputs to find a better structure of DAG for the decision policy. 
\begin{equation}
    f_t^{ij}(\bm{h}_t^{l}, \bm{h}_t^{r}, u) = \sigma(u^Ttanh(\bm{W_l}\bm{h}_t^{l} + \bm{W_r}\bm{h}_t^{r})),
\end{equation}
where $\bm{h}_t^{l}$ and $\bm{h}_t^{r}$ are agents' higher-level representations from two encoder outputs,  $\bm{W_l},\bm{W_r} \in \mathbb{R}^{d_h \times d_e}, u \in \mathbb{R}^{d_h \times 1}$ are trainable parameters, and $d_h, d_e$ are the hidden dimension and encoder output dimension, respectively. 

Moreover, the logistic sigmoid function $\sigma(\cdot)$ generates the probability for constructing the Bernoulli distribution from which  
the binary adjacency matrix $A$ is sampled. 
The binary adjacency matrix forms a directed graph corresponding to the ACG $\mathcal{G}$.
Here, we denote this graph generation process as $\mathcal{G} \sim \rho_{\varphi}(\bm{h})$.

\paragraph{\textbf{Parameter Setting.}}
In the~\namegg, the attention head in the GAT encoder is set to 8, the stacked attentional layers are set to 4, and the hidden units in the MLP is set to 64. 
In the~\namefp, the actor critic architecture is adopted. The recurrent layer comprised of a GRU with a 64-dimensional hidden state, with a fully-connected layer before and after, is used as the actor. The critic is a two-layer MLP with the ReLu activation.

\subsection{Algorithm Description}

The main procedures are summarized in
Algorithm~\ref{algo:nn}, where ${\nabla _\varphi }L (\varphi )$, ${\eta _{{\theta ^i}}}({\theta ^i})$,
and ${\nabla _\phi }\mathcal{L}(\phi )$ are optimized.
Our ultimate goal is to obtain the~\namefp~ $\{\pi^i\}_{i=1}^n$.
The~\namegg~$\rho$ is an intermediate
used to access 
an excellent graph structure in guiding 
the decision-making sequence among agents to achieve a high degree of multi-agent coordination. 
The graph generator is a pluggable module that can be replaced by other algorithms for solving the DAGs. Note
that DAGs are necessary because we need an execution structure that can determine a 
clear sequence of the actions, and therefore it should be directed and not circular. The policy solver 
is a universal module, which, in general,
one can choose from a diverse set of cooperative MARL algorithms~\cite{lowe2017maddpg,yu2021mappo,iqbal2019maac}.

\section{ Experiments}
We evaluate the effectiveness of our algorithm  on three different environments: Collaborative Gaussian Squeeze\footnote{The MGS environment is at \url{https://github.com/Sonkyunghwan/QTRAN}}~\cite{son2019qtran}, Cooperation Navigation\footnote{The code is at \url{https://github.com/openai/multiagent-particle-envs}}~\cite{lowe2017maddpg}, and Google Research Football\footnote{The code is at \url{https://github.com/google-research/football}}~\cite{kurach2020google}.

\begin{algorithm}[ht!]
\caption{The optimization of GCS.}
\label{algo:nn}
\textbf{Ensure} ~\namefp~$\pi^i$ and~\namegg~$\rho$;\\
\textbf{Initialize}~$\epsilon,\gamma,M,C$, and the replay buffer $\mathcal{D}\leftarrow \emptyset$;\\
\textbf{Initialize}~the parameters $\theta^i,\phi^i$ 
for the~\namefp~networks, where $i=1,...,n$, and $\varphi$
for the~\namegg~network;\\
\textbf{Initialize}~the target networks $\theta^{i^-}=\theta^i$ and $\phi^{i^-}=\phi^i$; \\
\For {\rm each episode}{
Initial state $\leftarrow \{o_0^i\}_{i=1}^n$; // drop $i$ as $\bm{o_0}$ for clarity; \\
\textbf{Initialize} $h_0^{(1)},...,h_0^{(n)}$ for RNN states;\\
\For{\rm each timestep $t$}{
Get $\mathcal{G}_t \sim \rho(\cdot | \bm{o_t})$; \\
Get the topological order $A_t=f(\mathcal{G}_t)$; \\
\For{\rm $i$ in order $A_t$}{
augment the observations as $\hat o_t^i = (o_t^i, u_{t-1}^i, u_t^{pa(i)}) $;\\
sample the action with $\epsilon$-greedy from $Q_t^{\pi^i}(\hat o_t^i, u_t^i;\theta^i)$;\\
Update RNN state $h_{t}^{(1)},...,h_{t}^{(n)}$ $\leftarrow$ $h_{t-1}^{(1)},...,h_{t-1}^{(n)}$ ;\\
Receive reward {${\bm{r}_t}$} and observe next state ${\bf{o}}_{t+1}$; \\
Add transition {$\{{\bm{o}}_t, {\bm{u}}_t, {\bm{r}}_t, {\bm{o}}_{t+1}\}$}
into $\mathcal{D}$;
}
}

\If{ $episodes > M$}{
Sample a minibatch {$\mathcal{B}=\{{\bm{o}}_j, {\bm{u}}_j, {\bm{r}}_j, {\bm{o}}_{j+1}\}^M_{j=0}\sim \mathcal{D}$}
; \\
Update the policy $\pi^i$ using $\mathcal{B}$ and (\ref{eq:policy});\\
Update the graph generator $\rho$ using $\mathcal{B}$ and ${\nabla _\varphi }L (\varphi )$;\\
Update $\phi^i$ using~(\ref{eq:critic});\\
Every $C$ steps reset $\theta_i^-=\theta_i$ for $i=1,...,n$;\\
Update the Lagrange penalty $\xi$ and the multipliers $\bm{\lambda}_1, \bm{\lambda}_2$;
}
}
\end{algorithm}



%% file: 4-experiment.tex
\subsection{Experimental Setting}

\label{sec:setting}
\paragraph{\textbf{Collaborative Gaussian Squeeze (CGS)}}

As an extension of Multi-domain Gaussian Squeeze (MGS)~\cite{son2019qtran}, Collaborative Gaussian Squeeze is a challenging environment for evaluating coordination.
In MGS, there exist $K$ domains $[({\mu _1},{\sigma _1}),...,({\mu _K},{\sigma _K})]$ in the system. 
The system contains $N$ agents, and each agent $i$ can take actions $a_i$ within range of 
$\{-10,-9,..,0,1,..,8,9,10\}$. 
The prior $s_i \in [0, 0.2]$ given by the environment represents the unit-level resource for each agent $i$. The total amount of resources mobilized by all agents is 
denoted as $f(\bm{u}) = \sum\nolimits_i s_i \times a_i$.
The goal is to maximize the joint reward $G(\bm{u}) = \sum\nolimits_{k = 1}^K {f(\bm{u}){e^{ - {{(f(\bm{u})- {\mu _k})}^2}/{\sigma _k}^2}}}$.
In our settings, we modify the original MGS to a collaborative task. 
We use two domains $[(\mu_1=5, \sigma_1=1.25),(\mu_2=-5, \sigma_2=1.25)]$.
The computation of the joint reward is as follows:
\begin{equation}
    G(\bm{u}) = f(\bm{u}){e^{ - {{(f(\bm{u}) - {\mu _1})}^2}/{\sigma _1}^2}} - f(\bm{u}){e^{ - {{(f(\bm{u}) - {\mu _2})}^2}/{\sigma _2}^2}}.
\end{equation}

\begin{figure}[ht]
    \centering
    \subfloat[CGS]{
	  \label{fig:exp-GS}
       \includegraphics[width=0.3\linewidth]{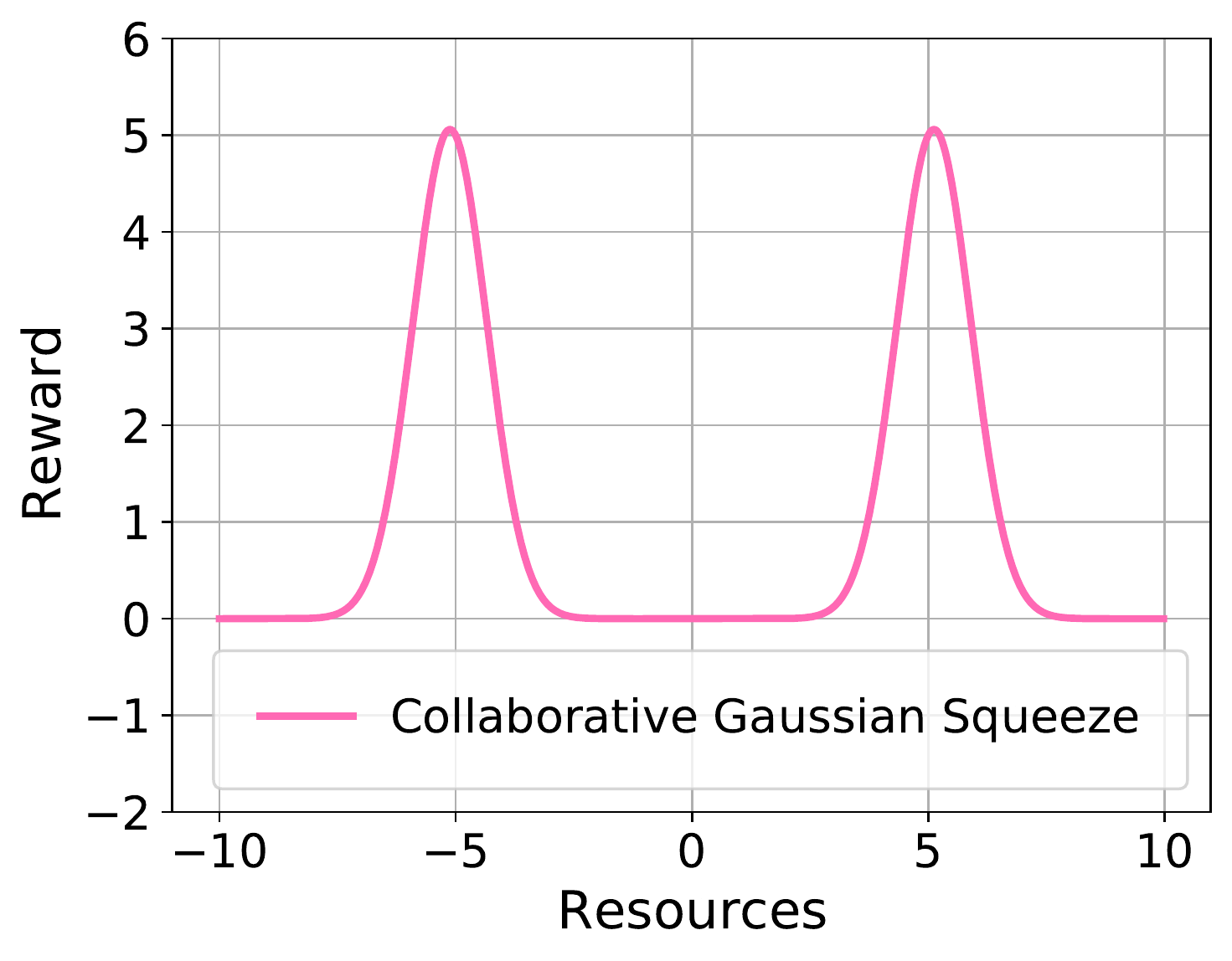}}
	\subfloat[CN]{
	  \label{fig:exp-Navigation}
      \includegraphics[width=0.3\linewidth]{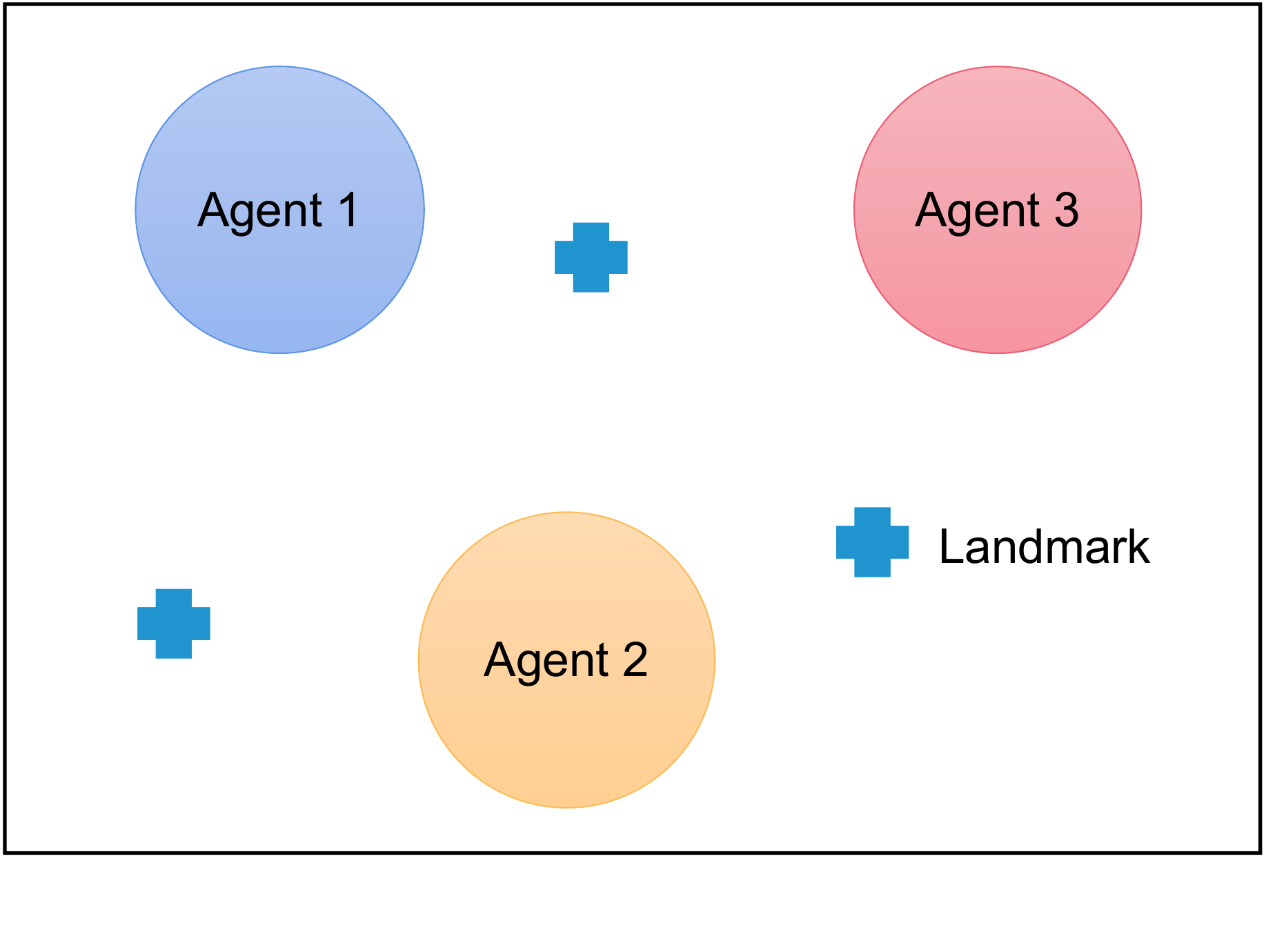}} 
    \subfloat[GRF]{
	  \label{fig:exp-Football}
      \includegraphics[width=0.3\linewidth]{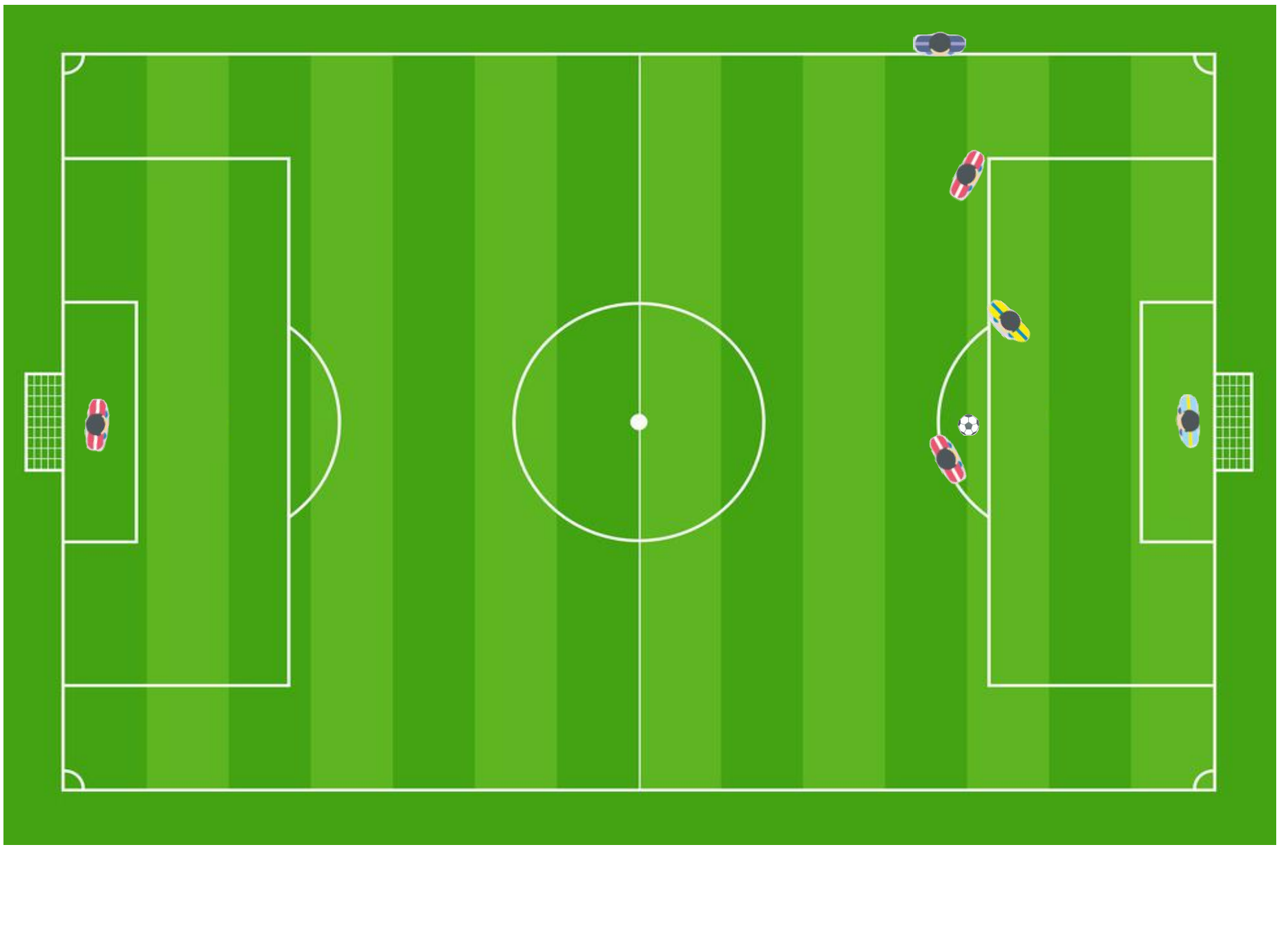}}
	  
	  \caption{The schematics of our experimental environments. 
	  }
	  \label{fig:env}
\end{figure}

Figure~\ref{fig:exp-GS} shows the reward curves for this setting. According to the above definition, the reward is maximized 
when the resources of all the agents reach  $\mu_1$ or $\mu_2$. 
In this environment, the social welfare depends on the intensity of collaboration.


\paragraph{\textbf{Cooperative Navigation (CN)}}  Cooperative Navigation is a classic scenario implemented in the multi-agent particle world. This scenario has $n$ agents and $n$ landmarks, which are initialized with random locations at the beginning of each episode. The objective of the agents is to cooperate to cover all landmarks by controlling their velocities with directions. The action set $\mathcal{A}$ includes five actions: \texttt{[up, down, left, right, stop]}. 
Each agent can only observe its velocity, position, and displacement from other agents and the landmarks. 
The shared reward is the negative sum of displacements between each landmark and its nearest agent. 
Agents must also avoid collisions, since each agent is penalized with a `$-1$' shared reward for every collision with other agents. 
We set the length of each episode as 25 time-steps. Therefore,  the agents have to learn to navigate toward the landmarks cooperatively 
to cover all positions quickly and accurately.
Figure~\ref{fig:exp-Navigation} shows a classic scenario in Cooperative Navigation with $N=3$.

\paragraph{\textbf{Google Research Football (GRF)} }
GRF is a realistically complicated and dynamic simulation environment without any clearly defined behavior abstraction, which is a suitable testbed for studying multi-agent decision making and coordination.
In GRF, we use the \textit{Floats} wrappers to represent the state. The \textit{Floats} representation contains a 115-dimensional vector that summarizes the information, such as the ball position and possession, coordinates of all players, and the game state. 
Each player has 19 actions to control, including the standard move actions and different ball-kicking techniques.
The rewards include the $SCORING$ reward $(-1,+1)$ and  the $CHECKPOINT$ reward, which is the shaped reward that specifically addresses the sparsity of $SCORING$. Detailed descriptions are shown in Appendix~\ref{detail:football}.

\paragraph{\textbf{Baselines}}
We compared our results with several baselines as follows. VDN and QMIX are  state-of-the-art value factorization approaches that follow the regime of centralized training and decentralized execution, belonging to the class of fully decentralized control policies, 
with which it is difficult to obtain coordinated behaviours.
DCG uses fully connected graphs for belief propagation, which only allows the message passing of paired agents.
DGN aims at learning abstract representations to make simultaneous decisions.

\begin{itemize}
    \item VDN~\cite{sunehag2018vdn}: Value Decomposition Network (VDN) imposes the structural constraint of  additivity in the factorization, which represents $Q_{tot}$ as a sum of individual Q-values.
    \item QMIX~\cite{rashid2018qmix}: This was proposed to overcome the limitation that VDN uses the linear decomposition and ignores any extra state information available during training. QMIX enforces  $Q_{tot}$ to be monotonic in the individual Q-values $Q^i$.
    \item DCG~\cite{bohmer2020dcg}: Deep Coordination Graph (DCG) factorizes the joint value function of all agents according to a coordination graph into payoffs between pairs of agents, which coordinates the actions between agents explicitly.
    \item DGN~\cite{jiang2018dgn}: {DGN relies on a graph convolutional network to model the relation representations, implicitly modeling the action coordination.}
\end{itemize}

%% file: 4-exp-performance.tex
\subsection{Main Results}
Here we report the experimental results from the setup described in Section~\ref{sec:setting}. Performance validation indicates the superiority of introducing ACG to  multi-agent systems.



\begin{figure}[h!]
\begin{center}
\includegraphics[width=0.66\linewidth]{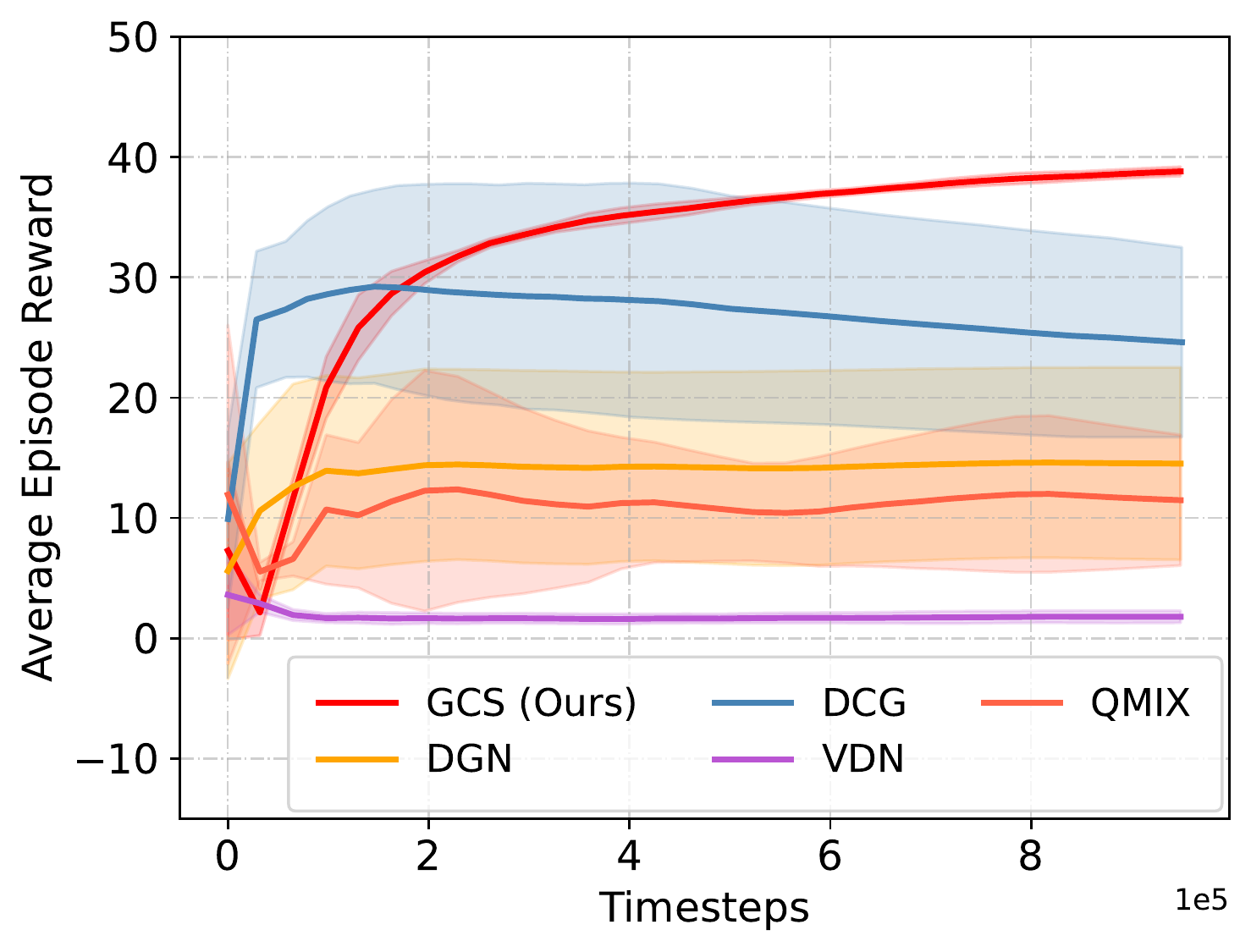}
\end{center}
\caption{Average episode rewards compared with  baselines on Collaborative Gaussian Squeeze.}
\label{fig:CGS_per}
\end{figure}

\paragraph{\textbf{Collaborative Gaussian Squeeze}}
In this game, there are 10 agents, and the maximum episode length is also set to 10.
To emphasize the feasibility and effectiveness of our proposed framework, we first conduct the experiment on CGS. We report the average episode rewards over 10 random runs, shown in Figure~\ref{fig:CGS_per}.
Our proposed algorithm GCS outperforms the baseline methods by a large margin. 
It can be seen that our algorithm  handles the collaborative problem well; the action coordination graph facilitates  behavior learning to promote cooperation.
Next, we will verify the effectiveness of our algorithm on more complicated environments.

\vspace{-10pt}

\begin{figure}[h!]
    \centering
	\subfloat[Four agents]{
	  \label{fig:exp-cop4}
       \includegraphics[width=0.48\linewidth]{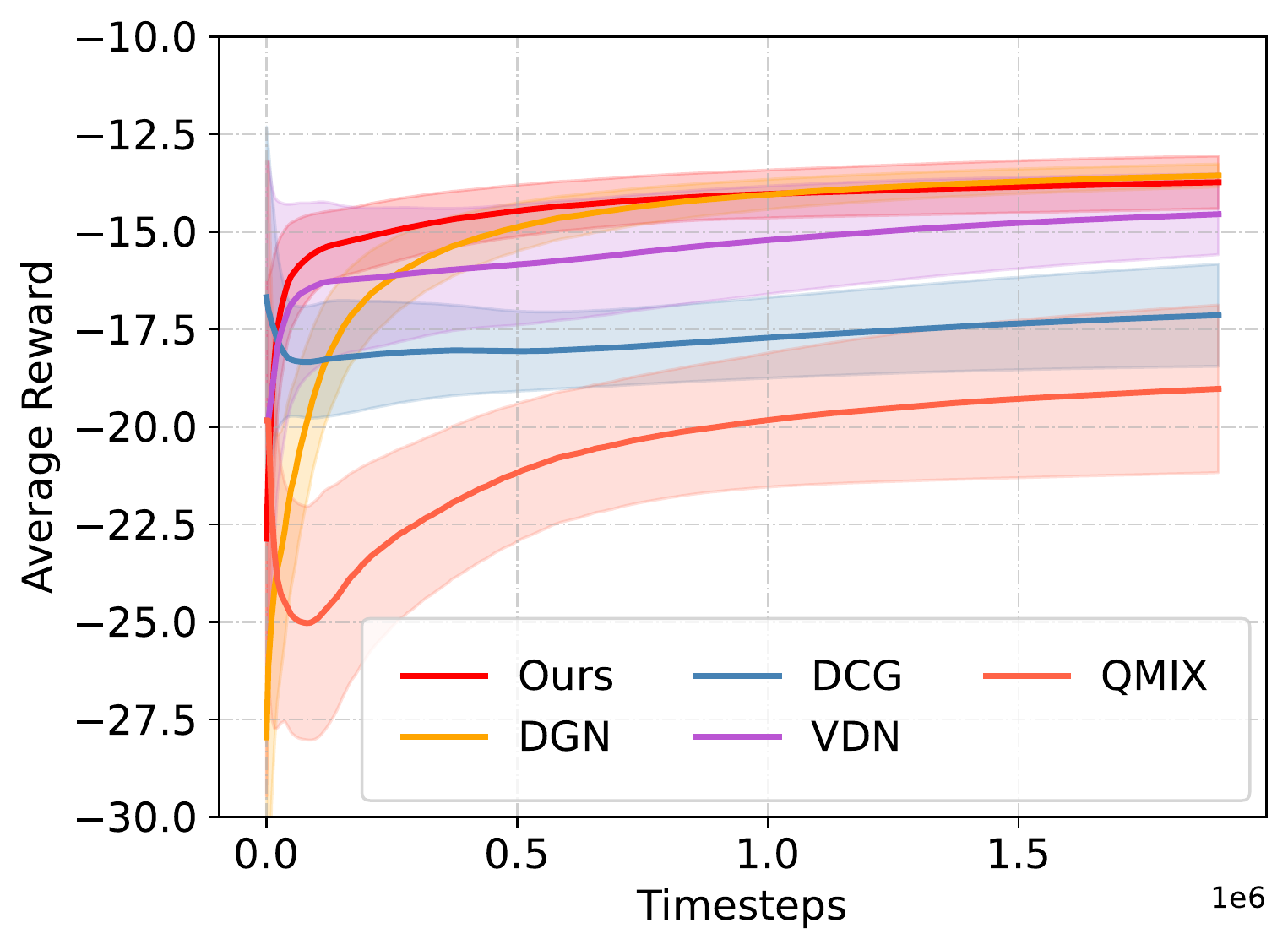}}
    \subfloat[Six agents]{
	  \label{fig:exp-cop6}
        \includegraphics[width=0.48\linewidth]{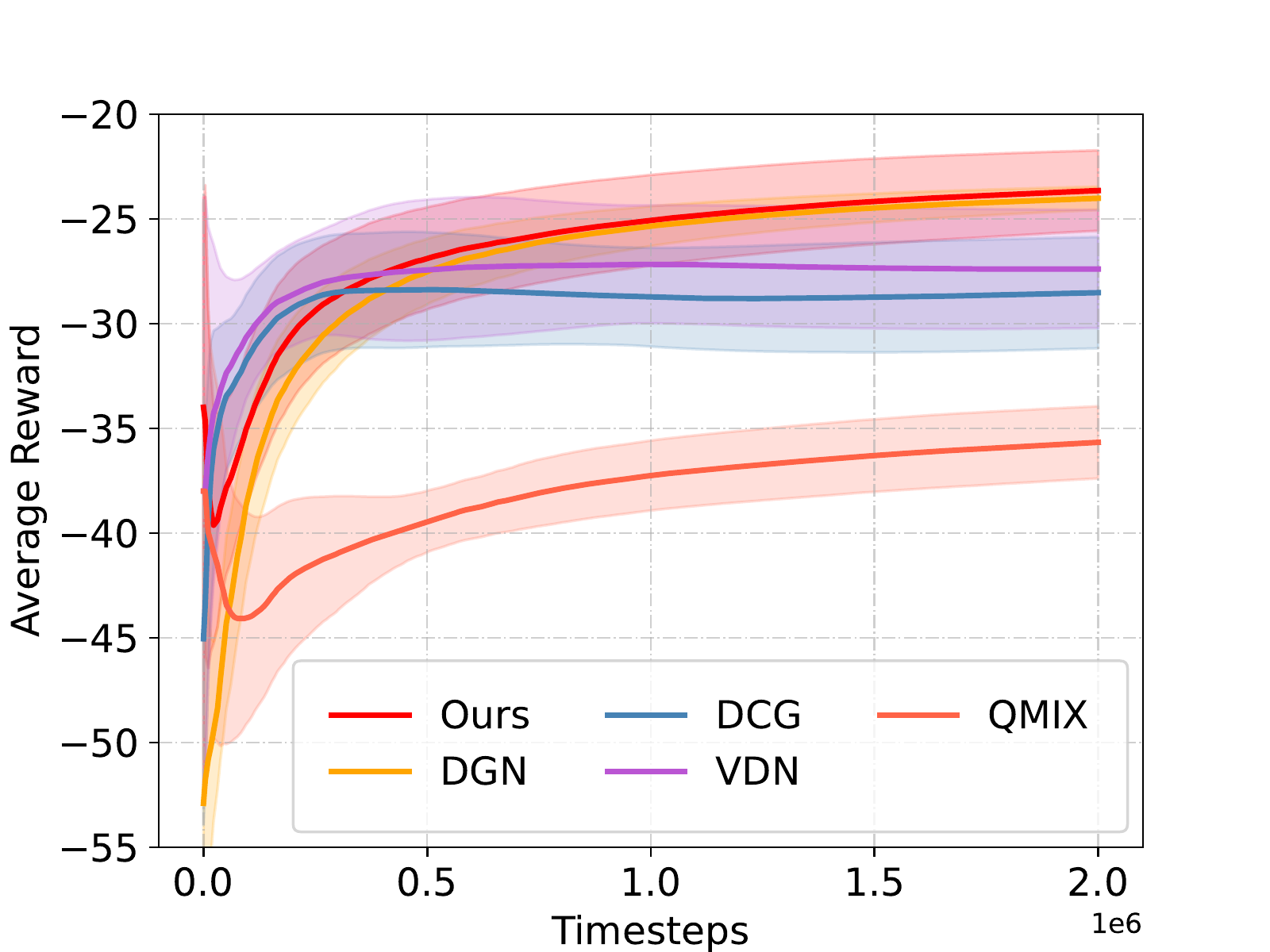}}\\
	  \caption{Average episode rewards for four agents and six agents in Cooperative Navigation.
	  }
	  \label{fig:exp-2}
\end{figure}

\paragraph{\textbf{Cooperative Navigation}}
As shown in Figure~\ref{fig:exp-Navigation}, Cooperative Navigation is a fully cooperative environment in which $N$ agents (circles) must
cooperate to reach $N$ landmarks (crosses) with as few collisions as possible.
We conduct experiments in  Cooperative Navigation with $N=4$ and with $N=6$.  Figures~\ref{fig:exp-cop4} and~\ref{fig:exp-cop6} show the learning curve comparisons for the two cases.
We report the average episode reward at a training step, averaged over 10 independent running seeds. 

First, our algorithm outperforms  most baseline algorithms by giving higher converged rewards. We consider that our performance improvement results from the action coordination graph representing the action dependency for better coordination.
Moreover, our algorithm converges fast during
training, which is possibly because the hierarchical decision policies can efficiently induce coordination among agents in this cooperative setting. 
In addition, our algorithm can achieve a lower variance than those baselines, which indicates that the learned action coordination graph can reduce the uncertainty in decision making to facilitate cooperative behaviors among agents.

In contrast, VDN and QMIX take actions simultaneously without considering the action dependency among agents. They are faster during training, but that is of no benefit in inducing cooperation among agents. 
Additionally, DCG exhibits mediocre performance in this task.
We believe that DCG considers only  pairwise relationships between agents, which may disturb the overall balance in the system.
In this case, DGN shows good performance consistent with ours, which shows that implicit action coordination modeling is also effective in pure cooperative settings.


\begin{figure}[h!]
    \centering
	\subfloat[Current state]{
	  \label{fig:visualize-1}
       \includegraphics[width=0.28\linewidth]{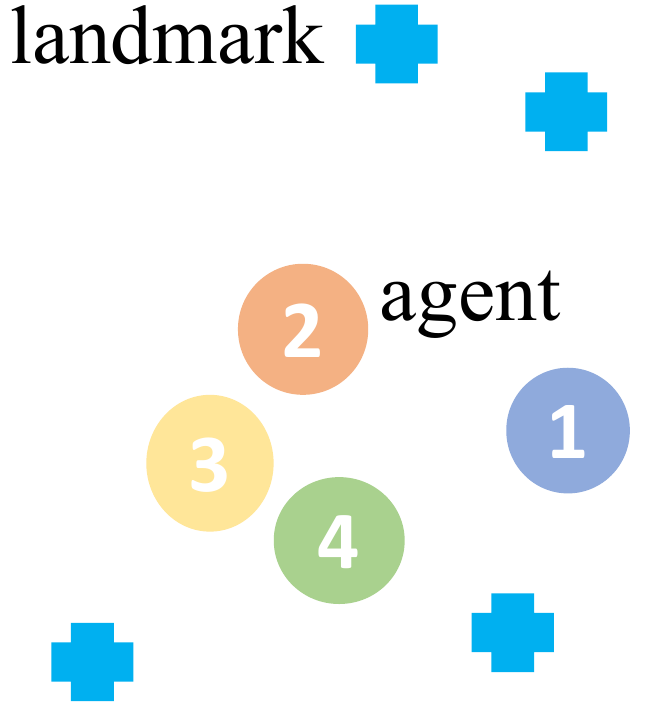}}
    \subfloat[Topology]{
        \label{fig:visualize-2}
        \includegraphics[width=0.36\linewidth]{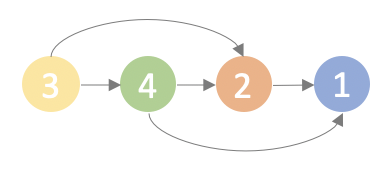}
    }
    \subfloat[Goals and actions]{
	  \label{fig:visualize-3}
        \includegraphics[width=0.28\linewidth]{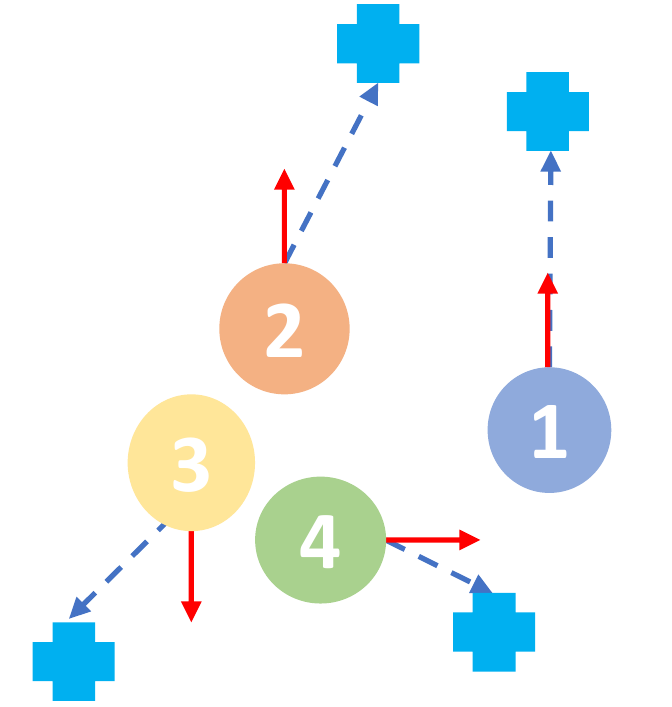}}\\
    
	\caption{Illustration of the effects of the action coordination graph. (a) Current state. (b) Topological structure of the learned ACG. (c) The intention of agents according to the ACG. The dashed lines represent the targets, and the solid lines represent the actions.}
	\label{fig:effect_acg}
	\vspace{-10pt}
\end{figure}

In order to clearly show the meaningful effect of the action coordination graph in the decision-making process, we give a visualization at a time step in an episode of Cooperative Navigation, as shown in Figure~\ref{fig:effect_acg}. 
Figure~\ref{fig:visualize-2} shows the topological structure of the learned ACG, and it suggests that the decision dependency of the agent is [3, 4, 2, 1].
The parent sets of the four agents are denoted as $pa(1)=\{4,2\}$, $pa(2)=\{3,4\}$, $pa(3)=\emptyset$, and $pa(4)=\{3\}$, respectively.
First, agent 3 decides to move to the bottom-left landmark, then agent 4 takes the best response and decides to move to the bottom-right in order to avoid conflict with agent 3. After agent 2 knows the decisions of {the previous two} agents, it chooses the closer upper-left as its target instead of the bottom-right. Finally, agent 1 moves after observing the decisions of agents 2 and 4. This visualization 
shows how agents' joint actions deriving from the ACG representing the underlying decision dependencies achieves efficiency.

\begin{figure*}[ht!]
    \centering
	\subfloat[3 vs. 2]{
	  \label{fig:exp-football-1}
       \includegraphics[width=0.3\linewidth]{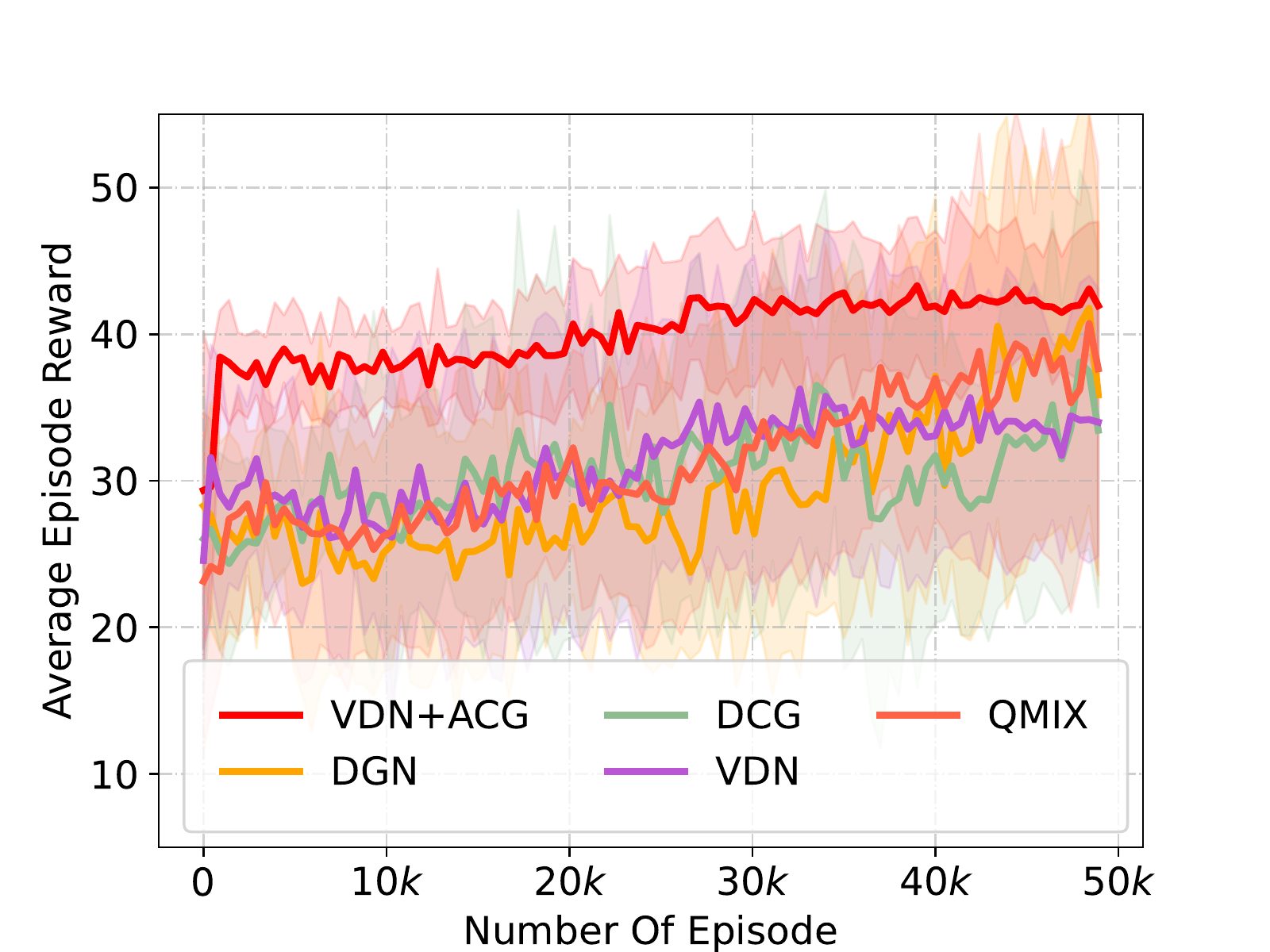}}
    \subfloat[3 vs. 6]{
	  \label{fig:exp-football-2}
       \includegraphics[width=0.3\linewidth]{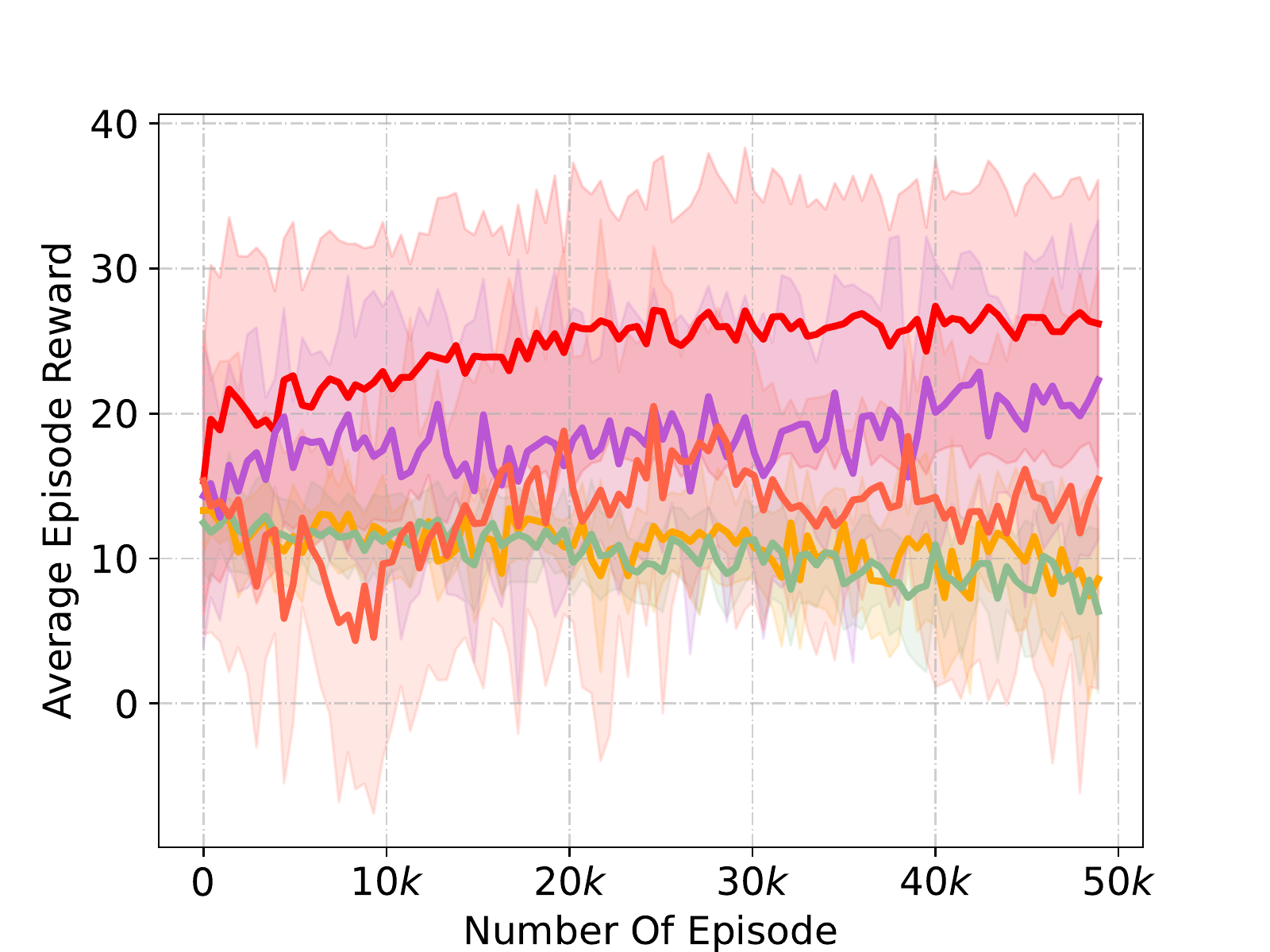}}
    \subfloat[5 vs. 5]{
	  \label{fig:exp-football-3}
        \includegraphics[width=0.3\linewidth]{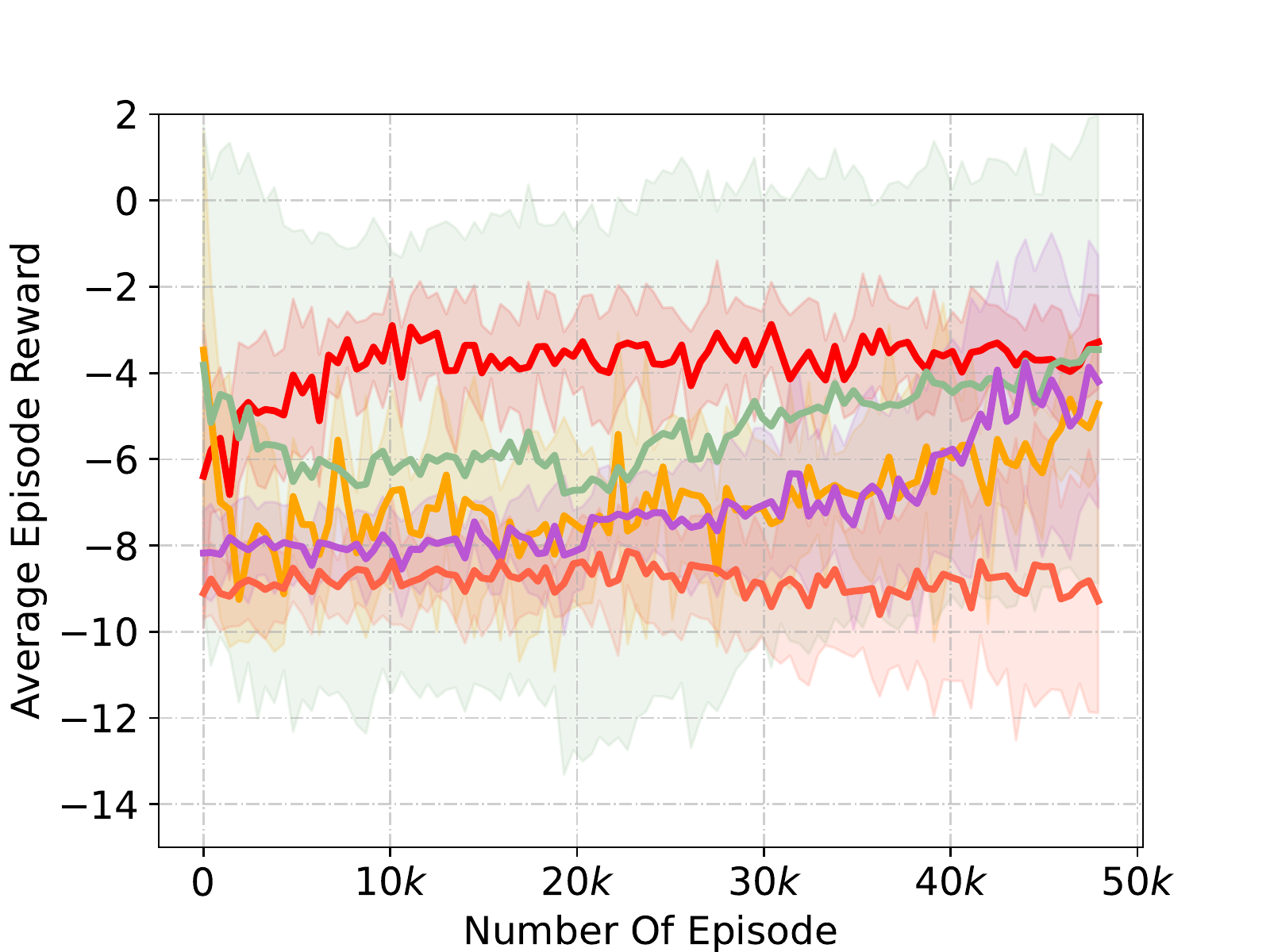}}
	  \caption{Average episode reward vs. training steps  for comparisons with the baselines on Google Research Football.
	  }
	  \label{fig:exp-2-football}
\end{figure*}

\paragraph{\textbf{Google Research Football (GRF)}}
To evaluate our method in complicated and dynamic environments, we conduct several experiments on GRF, as shown in Figure~\ref{fig:exp-2-football}. 
In the 3-vs-2 scenario, three of our players try to score from the edge of the box, and the opponent team contains one defender and one keeper. 
In the 3-vs-6 scenario, there are six opponent players on the pitch to play against three of our players.
In the 5-vs-5 scenario, each team has a keeper, an offensive player, and three defenders.
Here, we report the average episode reward at a training step for each scenario, averaged over 10 independent running seeds.

As can be seen in Figure~\ref{fig:exp-2-football}, our algorithm always obtains higher rewards than
all the baselines in the different scenarios of GRF. This indicates that our method is quite general in complicated and dynamic environments.
Moreover, this performance improvement in GRF demonstrates that our approach is good at effectively handling stochasticity and sparse rewards. This is because the learned ACG with the decision dependency is an efficient way to mitigate  uncertainty and induce cooperation among agents. 
Taking the 3-vs-6 scenario for further demonstration, the training curve of QMIX fluctuates and is unstable, indicating this method's inability to adapt to the dynamically complicated scenario with multiple opponent players. Here, DCG shows a trend of non-convergence, but our algorithm steadily rises to converge and obtains the highest reward, which exhibits the modeling supremacy of our approach for handling complicated tasks.



%% file: 4-exp-hierachy.tex

\subsection{Results on DAG Depth}

\begin{figure}[H]
\begin{center}
\includegraphics[width=0.6\linewidth]{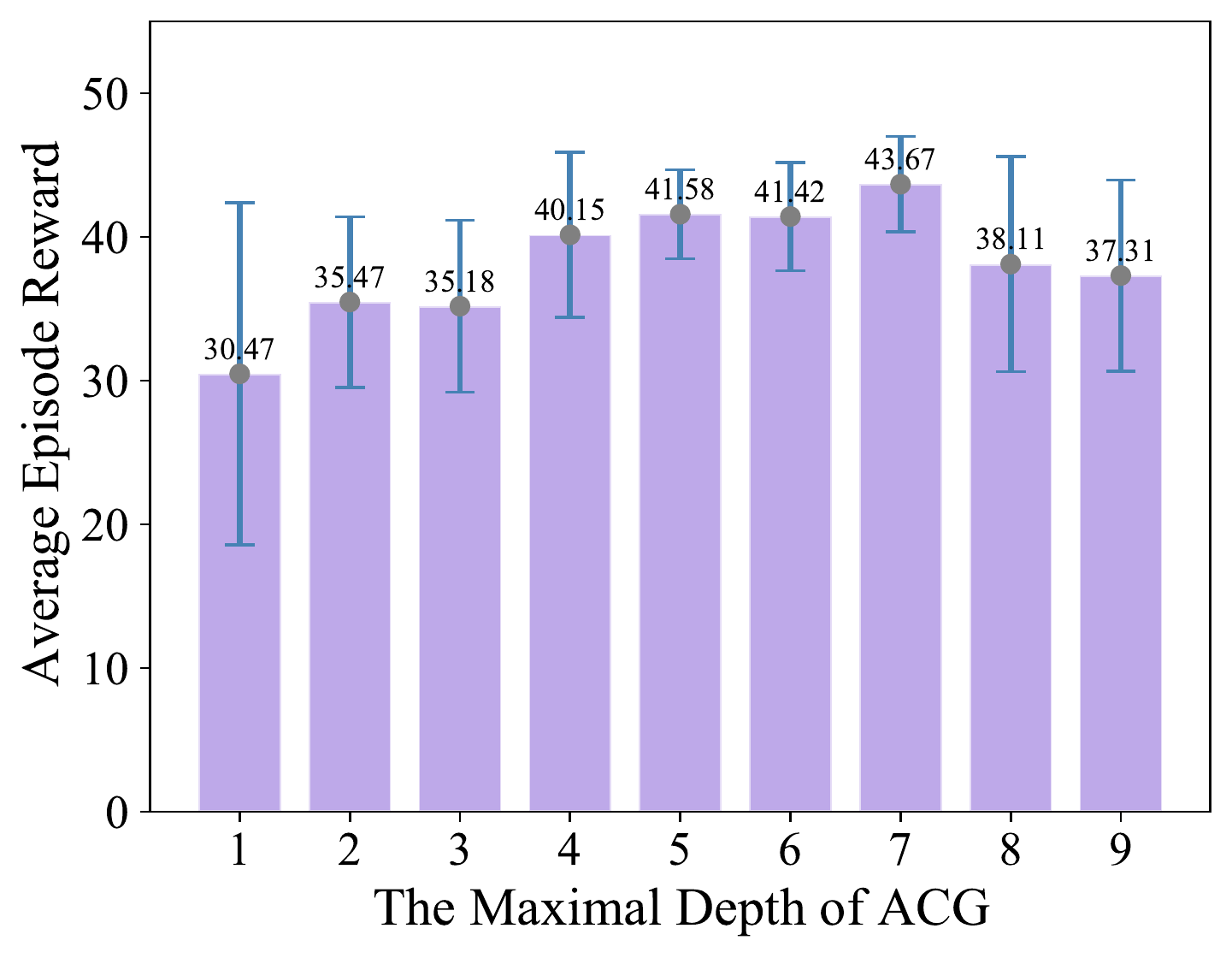}
\end{center}
\caption{Effect of different DAG depth constraints on Collaborative Gaussian Squeeze.}
\label{fig:hire_comp}
\end{figure}

We observe that the inference efficiency and the performance gains are inversely affected by the ACG's depth.
Therefore, we aim to find the suitable depth $k$ that best balances the tradeoff.
As shown in Figure~\ref{fig:hire_comp}, to validate the impact of the depth, we test our method on the Collaborative Gaussian Squeeze with different depth sizes of the learned ACG.
In this figure, the horizontal axis is the depth, and the vertical axis is the testing episode reward averaged over five seeds.
We test $1000$ episodes for each seed and obtain an average episode reward.


\vspace{-1em}

As the depth of the ACG increases, the training time will increase correspondingly. However, the performance growth will gradually slow down, and  performance degradation may even occur. In this case, $k=5$ is the optimal depth that balances the computational burden and the performance gains. 
As for the reason for the performance degradation with $k=8$ and $k=9$, we speculate that as the hierarchy of action dependency deepens, the complexity of the hypothesis space for the inference will increase, and it becomes harder to learn the optimal policy.
In summary, a higher dependency level of the graph structure can provide more decision information to promote coordination and facilitate performance. However, this higher dependency level leads to a lower efficiency of inference, as the leaf node on the ACG needs to wait for all the parent nodes' decisions before it makes its own decision.



%% file: 4-exp-robust.tex
\subsection{Results on Dropping Edges}
In this section, we verify the stability of the learned ACG in our proposed algorithm. Given a  trained model with $depth=5$ on Collaborative Gaussian Squeeze, we evaluate 1000 episodes and count the average number of edges of ACG, denoted as $edges_{num} \approx 28$. 
A fair comparison requires
the same depth and number of edges during training and evaluation.
Therefore, we generate a fixed DAG structure $\mathcal{G}_{\{5,28\}}$, see Appendix~\ref{app:DAG_struct}, whose depth is $5$ and whose number of edges is $28$, as the baseline to compare with our algorithm.


\begin{figure}[H]
    \centering
	\subfloat[Baseline $\mathcal{G}_{\{5,28\}}$ ]{
	  \label{fig:exp-robust1}
       \includegraphics[width=0.48\linewidth]{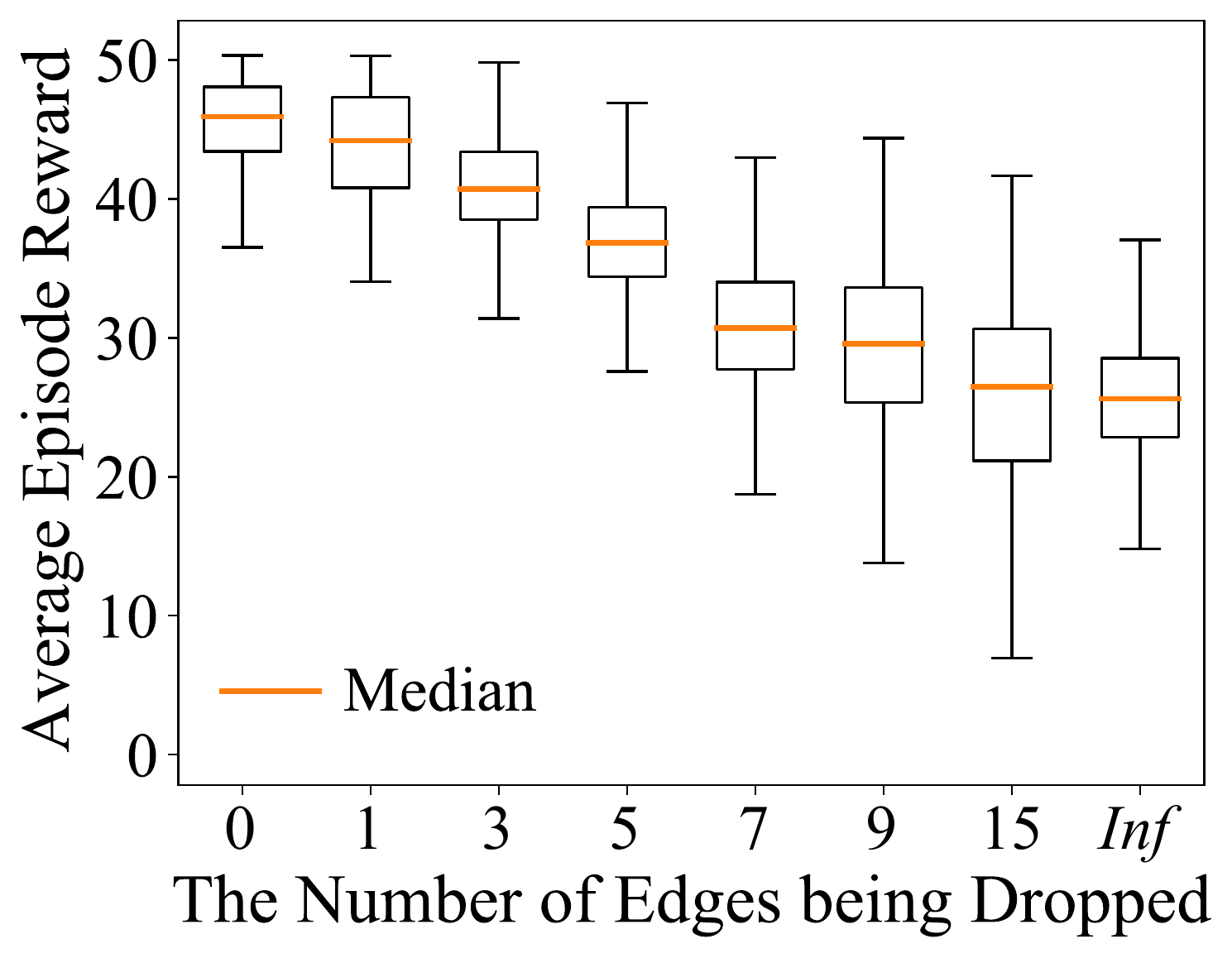}}
    \subfloat[GCS (ours)]{
	  \label{fig:exp-robust2}
        \includegraphics[width=0.48\linewidth]{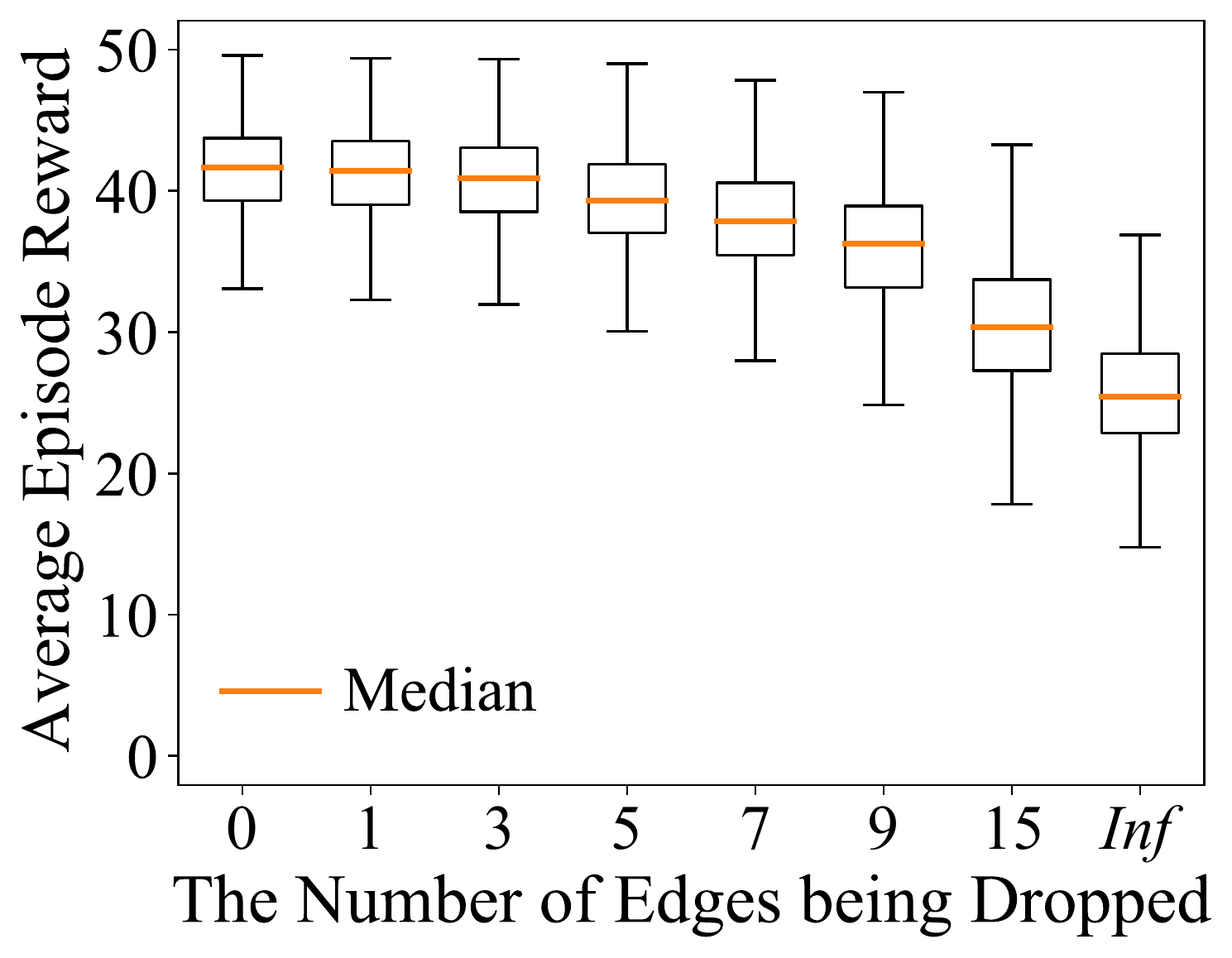}}\\
	  \caption{Box plots showing the distribution of the testing episode rewards in Collaborative Gaussian Squeeze.}
	  \label{fig:robust_comp}
\end{figure}

In Figure~\ref{fig:robust_comp}, the horizontal axis represents the number of edges  dropped from the generated or fixed graph structure, denoted as $\#drop$, and the vertical axis is the testing average episode reward over 1000 episodes. $Inf$ denotes dropping all the edges. The box plots visually show the distribution of the testing episode rewards and skewness by displaying the data quartiles. From the overall trend of Figure~\ref{fig:robust_comp}, we can observe that the data quartiles of the baseline reduce faster and change more drastically than our algorithm when the disturbance of edge dropping increases. This demonstrates that our algorithm has better stability.
Moreover, in Table~\ref{tab:robust_mean},
our algorithm outperforms the baseline with higher mean rewards in most cases, which demonstrates the power of the ACG learned by our model to promote the coordination among agents reliably and stably even when confronted with disturbances of different intensities. 
It is worth noting that to guarantee stability, a slight performance loss may occur. It will be an interesting future research direction to study the stability--performance trade-off.

\begin{table}[H]
\caption{Mean value of testing episode rewards when dropping  corresponding numbers of edges.}
\small
\setlength{\tabcolsep}{1.3mm}{
\begin{tabular}{ccccccccc@{}}
\toprule
\diagbox{\textbf{{Methods}}}{{\textbf{\#drop}}}  &  0 &  1    & 3    & 5    & 7    & 9    & 15   & inf  \\ \hline
\textbf{The baseline}  & 45.3 & 44.0 & 40.7 & 36.9 & 30.8 & 29.3 & 25.8 & 26.2 \\
\textbf{GCS (ours)}    & 41.6 & 41.3 & 40.9 & 39.4 & 37.9 & 36.0 & 30.5 & 26.1 \\ \bottomrule
\end{tabular}}

\label{tab:robust_mean}
\end{table}

In summary, comparisons with VDN, QMIX, and DCG on three environments demonstrate that our algorithm achieves better performance, stronger stability, and more powerful modeling capability for handling dynamically complicated tasks than any of these methods. 
Moreover, the proposed DAG depth constraint provides an insightful view on balancing efficiency and performance.


%% file: 5-conclusion.tex
\section{Conclusions}
In this paper, we introduce a novel~\namegg~ and~\namefp~in MARL to dynamically represent the underlying decision dependency structure and facilitate behavior learning, respectively. 
We propose the DAGness-constrained and DAG depth-constrained optimization to balance  training efficiency and  performance gains.
Extensive empirical experiments on Collaborative Gaussian Squeeze, Cooperative Navigation, and Google Research Football, as well as comparisons to baseline algorithms, demonstrate the superiority of our method.

Future research may consider
improving the limited performance by upgrading the~\namegg~model. We will also investigate an automatic mechanism for finding an appropriate depth for the action coordination graph.



%% file: 6-ack.tex
\section*{Acknowledgements}
Jingqing Ruan is supported in part by the Strategic Priority Research Program of the Chinese Academy of Sciences under Grant XDA27010404 and in part by the National Nature Science Foundation of China under Grant 62073324. Co-author Haifeng Zhang is supported in part by the Strategic Priority Research Program of the Chinese Academy of Sciences, Grant No. XDA27030401.

%% file: appendix.tex
\section{Detailed Proofs}
We provide the detailed proofs of the propositions in the following.
\subsection{Proof of Proposition \ref{pro:hierarchy}}
\label{proof:hierarchy}
\begin{proof}[proof]
Firstly, we prove that the entry $A^k_{ij}$ in $k$-th power of the adjacency matrix $A$ indicates the number of walks of length $k$ from node $v_i$ to $v_j$. Let $\mathcal{L}^l(i,j)$  denote the number of walks of length $l$ from node $v_i$ to $v_j$. When $l=1$, $A^1 = A, a_{ij}=\mathcal{L}^1(i,j)$. Let $b_{ij} =\mathcal{L}^k(i,j)$ denote the $ij^{th}$ entry of $A^k$. We have ${A_{ij}^{k + 1}} = A_{ij}{A_{ij}^k} = {a_{i1}}{b_{1j}} + {a_{i2}}{b_{2j}} + ... + {a_{in}}{b_{nj}} = \sum\nolimits_n {{a_{in}}{b_{nj}}}$, where ${\forall _m} \in [1,n],{a_{im}}{b_{mj}} = {\mathcal{L}^1}(i,m) \cdot {\mathcal{L}^k}(m,j) = {\mathcal{L}^{k + 1}}(i,j)$. So $A_{ij}^{k + 1}$ denote the number of walks of length $k+1$ from $v_i$ to $v_j$. 
Then:
\begin{equation}
\label{eq:hire_constr}
    \begin{array}{l}
{d}(A^k): = sum({A^k}) = \sum\nolimits_i {\sum\nolimits_j {A_{ij}^k = 0} } \\
{d}(A^{k - 1}): = sum({A^{k - 1}}) = \sum\nolimits_i {\sum\nolimits_j {A_{ij}^{k - 1} > 0} } 
\end{array}.
\end{equation}
The equations represent there is not a walk of length $k$ and there is at least one walk of length $k-1$ from $v_i$ to $v_j$ respectively, that is ${A^k} = O, {A^{k - 1}} \ne O$. We define the hierarchy of the DAGs as the longest path length. Therefore, the hierarchy of the DAGs is $k$ if $A$ is the Nilpotent Matrix of index $k$.
\end{proof}

\subsection{Proof of Proposition \ref{pro:grad_rho}}
\label{proof:grad_rho}
\begin{proof}
Let ${{{ \pi }^i}}$ represent the fixed policy for agent $i$ trained by the~\namefp. $\varphi$ is the parameter to be solved by the~\namegg~$\rho$.  $g(\cdot)$ and $d(\cdot)$ are the constraint function as shown in Equation~(\ref{eq:dag_constr}) and Equation~(\ref{eq:hire_constr}). 
${Q_{{{ \pi }^i}}}( \cdot )$ denotes the state action function.
The gradients is derived as follows.

\begin{equation*}
\begin{array}{l}
{\nabla _\varphi }L(\varphi ,{\lambda _1},{\lambda _2}) = {\nabla _\varphi }\sum\limits_s {{p_{{{ \pi }^i}}}(s)\sum\limits_W {\rho_{\varphi} (W|s)} \sum\limits_{{u^i}} { {{{ \pi }^i}({u^i}|s,W) \cdot {{Q}_{{{ \pi }^i}}}(s,{u^i})} } } \\
\left. { - {\lambda _1}g(W) - {\lambda _2}{d}(W^k)} \right]\\
 = {\nabla _\varphi }\sum\limits_s {{p_{{{ \pi }^i}}}(s)\sum\limits_W {\rho_{\varphi} (W|s)} \sum\limits_{{u^i}} {\left[ {{{ \pi }^i}({u^i}|{o^i},{u^{pa(i)}}) \cdot \left[ {{Q_{{{ \pi }^i}}}({o^i},{u^i})} \right.} \right.} } \\
\left. { - {\lambda _1}[tr({e^{W \circ W}}) - d] - {\lambda _2}sum({W^k})} \right]\\
 = {\mathbb{E}_{s\sim p,W\sim \rho_{\varphi} ( \cdot |s)}}\left[ {{\nabla _\varphi }\log \rho_{\varphi} (W|s)\sum\limits_{{u^i}} {{{ \pi }^i}({u^i}|{o^i},{u^{pa(i)}}) \cdot }  {{Q_{{{ \pi }^i}}}({o^i},{u^i})} }\right. \\
\left. { - {\lambda _1}{{({e^{W \circ W}})}^T} \cdot 2W - {\lambda _2}\sum\nolimits_{i,j} {{{\left[ {k{W^k}{W^{ - 1}}} \right]}_{ij}}} } \right]
\end{array}
\end{equation*}

\end{proof}

\section{Deatils of Google Research Football}
\label{detail:football}
\paragraph{\textbf{Observations}}
The environment exposes the raw observations as Table~\ref{tab:GRF-obs}.
We use the \textit{Simple115StateWrapper}\footnote{We refer the reader to:\url{https://github.com/google-research/football} for details of encoded information.} as the simplified representation of a game state encoded with 115 floats. 



\begin{table}[]
\caption{The main information and  detailed descriptions about the observations in GRF.}
\begin{tabular}{@{}ccl@{}}
\toprule
\textbf{Information}  & \multicolumn{2}{c}{\textbf{Descriptions}} \\ \midrule 
\multirow{5}{*}{Ball information}                                                                     & \multicolumn{2}{c}{position of ball}      \\
                                                                                          & \multicolumn{2}{c}{direction of ball}                   \\
                                                                                          & \multicolumn{2}{c}{rotation angles of ball}                 \\
                                                                                          & \multicolumn{2}{c}{owned team of ball}                 \\
                                                                                          & \multicolumn{2}{c}{owned player of ball}                 \\ \midrule
\multirow{6}{*}{Left team}                                                                & \multicolumn{2}{c}{position of players in left team}      \\
                                                                                          & \multicolumn{2}{c}{direction of players in left team}                   \\
                                                                                          & \multicolumn{2}{c}{tiredness factor of players}                 \\
                                                                                          & \multicolumn{2}{c}{ numbers of players with yellow card}                 \\
                                                                                          & \multicolumn{2}{c}{whether a player got a red card}                 \\
                                                                                          & \multicolumn{2}{c}{roles of players}                     \\ \midrule
\multirow{6}{*}{Right team}                                                                     & \multicolumn{2}{c}{position of players in right team}      \\
                                                                                          & \multicolumn{2}{c}{direction of players in right team}                   \\
                                                                                          & \multicolumn{2}{c}{tiredness factor of players}                 \\
                                                                                          & \multicolumn{2}{c}{numbers of players with yellow card}                 \\
                                                                                          & \multicolumn{2}{c}{whether a player got a red card}                 \\
                                                                                          & \multicolumn{2}{c}{roles of players}                     \\ \midrule
\multirow{3}{*}{\begin{tabular}[c]{@{}c@{}} Controlled player information\end{tabular}} & \multicolumn{2}{c}{controlled player index}      \\
                                                                                          & \multicolumn{2}{c}{designated player index}                   \\
                                                                                          & \multicolumn{2}{c}{active action}                 \\ \midrule
\multirow{3}{*}{Match state}                                                              & \multicolumn{2}{c}{goals of left and right teams}      \\
                                                                                          & \multicolumn{2}{c}{left steps}                   \\
                                                                                          & \multicolumn{2}{c}{current game mode}                 \\ \midrule
Screen                                                                                    & \multicolumn{2}{c}{rendered screen}                     \\ \bottomrule
\end{tabular}

\label{tab:GRF-obs}
\end{table}
\paragraph{\textbf{Actions}}
The number of actions available to an individual agent can be denoted as $|\mathcal{A}| = 19$.
The standard move actions (in $8$ directions) include $\mathcal{A}_{move} = \{Top, Bottom, Left, Right,Top-Left, Top-Right, Bottom-Left, Bottom-Right\}$.
Moreover, the actions represent different ways to kick the ball is 
$\mathcal{A}_{kick} = \{Short Pass, High Pass,$\\
$ Long Pass, Shot,Do-Nothing,Sliding,Dribble,Stop-Dribble,Sprint, $\\
$Stop-Moving, Stop-Sprint\}$.

\paragraph{\textbf{Rewards}}
The reward function mainly includes two parts. The first is $SCORING$, which corresponds to the natural reward where the team obtains $+1$ when scoring a goal and $-1$ when losing one to the opposing team. The second part is $CHECKPOINT$, which is proposed to address the issue of sparse rewards. It is encoded with  domain
knowledge by an additional auxiliary reward contribution. For example, we can increase the reward when the player owns the ball to boost passing the ball.

\section{The detailed structure of $\mathcal{G}_{\{5,28\}}$}
\label{app:DAG_struct}
The adjacency matrix of $\mathcal{G}_{\{5,28\}}$ is shown as follows.
\begin{equation*}
     \left[ {\begin{array}{*{20}{c}}
{0{\quad}1\quad0\quad1\quad0\quad1\quad1\quad0\quad1\quad0}\\
{0\quad0\quad0\quad1\quad0\quad1\quad1\quad0\quad1\quad0}\\
{0\quad1\quad0\quad1\quad0\quad1\quad1\quad0\quad1\quad0}\\
{0\quad0\quad0\quad0\quad0\quad0\quad0\quad0\quad0\quad0}\\
{0\quad1\quad0\quad1\quad0\quad0\quad0\quad0\quad1\quad0}\\
{0\quad0\quad0\quad1\quad0\quad0\quad0\quad0\quad1\quad0}\\
{0\quad0\quad0\quad0\quad0\quad0\quad0\quad0\quad0\quad0}\\
{0\quad1\quad0\quad1\quad0\quad1\quad1\quad0\quad1\quad0}\\
{0\quad0\quad0\quad0\quad0\quad0\quad0\quad0\quad0\quad0}\\
{0\quad1\quad0\quad1\quad0\quad1\quad0\quad0\quad1\quad0}
\end{array}} \right]
\end{equation*}

\section{Additional Experimental Details}

We set discount factor $\gamma = 0.99$. The optimization is conducted using RMSprop with a learning rate of $5 \times 10^{-4}$ and $\alpha = 0.99$ with no weight decay. Exploration for action selection is performed during training, and each agent executes $\epsilon-greedy$ policy over its actions. $\epsilon$ is annealed from $0.2$ to $0.05$ over the first $50k$ time steps and is kept constant afterwards. 
 
 In addition, the information regarding computational resources used is Enterprise Linux Server with 96 CPU cores and 6 Tesla K80 GPU cores(12G memory).